\documentclass[11pt]{article}
\usepackage{amsmath,amsthm,latexsym,amssymb,amsfonts,epsfig,psfrag}

\makeatletter \@addtoreset{equation}{section} \makeatother

\def\be{\begin{equation}}
\def\ee{\end{equation}}
\def\ba{\begin{eqnarray}}
\def\ea{\end{eqnarray}}

\newcommand\nn{\nonumber}
\newcommand\q{\quad}
\def\Nl{{\mathchoice
{\setbox0=\hbox{$\displaystyle\rm N$}\hbox{\hbox to0pt
{\kern0.4\wd0\vrule height0.9\ht0\hss}\box0}}
{\setbox0=\hbox{$\textstyle\rm N$}\hbox{\hbox to0pt
{\kern0.4\wd0\vrule height0.9\ht0\hss}\box0}}
{\setbox0=\hbox{$\scriptstyle\rm N$}\hbox{\hbox to0pt
{\kern0.4\wd0\vrule height0.9\ht0\hss}\box0}}
{\setbox0=\hbox{$\scriptscriptstyle\rm N$}\hbox{\hbox to0pt
{\kern0.4\wd0\vrule height0.9\ht0\hss}\box0}}}}
\def\Zl{{\mathchoice
{\setbox0=\hbox{$\displaystyle\rm Z$}\hbox{\hbox to0pt
{\kern0.4\wd0\vrule height0.9\ht0\hss}\box0}}
{\setbox0=\hbox{$\textstyle\rm Z$}\hbox{\hbox to0pt
{\kern0.4\wd0\vrule height0.9\ht0\hss}\box0}}
{\setbox0=\hbox{$\scriptstyle\rm Z$}\hbox{\hbox to0pt
{\kern0.4\wd0\vrule height0.9\ht0\hss}\box0}}
{\setbox0=\hbox{$\scriptscriptstyle\rm Z$}\hbox{\hbox to0pt
{\kern0.4\wd0\vrule height0.9\ht0\hss}\box0}}}}
\def\Ql{{\mathchoice
{\setbox0=\hbox{$\displaystyle\rm Q$}\hbox{\hbox to0pt
{\kern0.4\wd0\vrule height0.9\ht0\hss}\box0}}
{\setbox0=\hbox{$\textstyle\rm Q$}\hbox{\hbox to0pt
{\kern0.4\wd0\vrule height0.9\ht0\hss}\box0}}
{\setbox0=\hbox{$\scriptstyle\rm Q$}\hbox{\hbox to0pt
{\kern0.4\wd0\vrule height0.9\ht0\hss}\box0}}
{\setbox0=\hbox{$\scriptscriptstyle\rm Q$}\hbox{\hbox to0pt
{\kern0.4\wd0\vrule height0.9\ht0\hss}\box0}}}}
\def\Rl{{\mathchoice
{\setbox0=\hbox{$\displaystyle\rm R$}\hbox{\hbox to0pt
{\kern0.4\wd0\vrule height0.9\ht0\hss}\box0}}
{\setbox0=\hbox{$\textstyle\rm R$}\hbox{\hbox to0pt
{\kern0.4\wd0\vrule height0.9\ht0\hss}\box0}}
{\setbox0=\hbox{$\scriptstyle\rm R$}\hbox{\hbox to0pt
{\kern0.4\wd0\vrule height0.9\ht0\hss}\box0}}
{\setbox0=\hbox{$\scriptscriptstyle\rm R$}\hbox{\hbox to0pt
{\kern0.4\wd0\vrule height0.9\ht0\hss}\box0}}}}
\def\Cl{{\mathchoice
{\setbox0=\hbox{$\displaystyle\rm C$}\hbox{\hbox to0pt
{\kern0.4\wd0\vrule height0.9\ht0\hss}\box0}}
{\setbox0=\hbox{$\textstyle\rm C$}\hbox{\hbox to0pt
{\kern0.4\wd0\vrule height0.9\ht0\hss}\box0}}
{\setbox0=\hbox{$\scriptstyle\rm C$}\hbox{\hbox to0pt
{\kern0.4\wd0\vrule height0.9\ht0\hss}\box0}}
{\setbox0=\hbox{$\scriptscriptstyle\rm C$}\hbox{\hbox to0pt
{\kern0.4\wd0\vrule height0.9\ht0\hss}\box0}}}}
\def\Hl{{\mathchoice
{\setbox0=\hbox{$\displaystyle\rm H$}\hbox{\hbox to0pt
{\kern0.4\wd0\vrule height0.9\ht0\hss}\box0}}
{\setbox0=\hbox{$\textstyle\rm H$}\hbox{\hbox to0pt
{\kern0.4\wd0\vrule height0.9\ht0\hss}\box0}}
{\setbox0=\hbox{$\scriptstyle\rm H$}\hbox{\hbox to0pt
{\kern0.4\wd0\vrule height0.9\ht0\hss}\box0}}
{\setbox0=\hbox{$\scriptscriptstyle\rm H$}\hbox{\hbox to0pt
{\kern0.4\wd0\vrule height0.9\ht0\hss}\box0}}}}
\def\Ol{{\mathchoice
{\setbox0=\hbox{$\displaystyle\rm O$}\hbox{\hbox to0pt
{\kern0.4\wd0\vrule height0.9\ht0\hss}\box0}}
{\setbox0=\hbox{$\textstyle\rm O$}\hbox{\hbox to0pt
{\kern0.4\wd0\vrule height0.9\ht0\hss}\box0}}
{\setbox0=\hbox{$\scriptstyle\rm O$}\hbox{\hbox to0pt
{\kern0.4\wd0\vrule height0.9\ht0\hss}\box0}}
{\setbox0=\hbox{$\scriptscriptstyle\rm O$}\hbox{\hbox to0pt
{\kern0.4\wd0\vrule height0.9\ht0\hss}\box0}}}}
\newcommand{\cd}{\mathcal D}

\newcommand{\cf}{\mathcal F}

\newcommand{\ck}{\mathcal K}
\newcommand{\cl}{\mathcal L}

\newcommand{\cp}{\mathcal P}

\newcommand{\calr}{\mathcal R}
\newcommand{\cs}{\mathcal S}

\newcommand{\cv}{\mathcal V}

\newcommand{\cz}{\mathcal Z}

\newcommand{\fe}{\mathfrak{e}}

       \newcommand{\Fraki}{\mathfrak{I}}

\newcommand{\fm}{\mathfrak{m}}

  \newcommand{\Fr}{\mathfrak{R}}

\newcommand{\dsty}{\displaystyle}

\title{Spin foam models with finite groups}
\author{Benjamin Bahr$^1$, Bianca Dittrich$^2$, James P.~Ryan$^2$\\
\small $^1$ DAMTP, University of Cambridge,\\
\small  Wilberforce Road, Cambridge CB3 0WA, UK\\
\small   $^2$ MPI f. Gravitational Physics, Albert Einstein Institute,\\
 \small Am M\"uhlenberg 1, D-14476 Potsdam, Germany
 }

\date{}

\begin{document}

\maketitle

\begin{abstract}
\noindent

Spin foam models, loop quantum gravity and group field theory  are discussed as quantum gravity candidate theories and usually involve a continuous Lie group. We advocate here to consider quantum gravity inspired models with finite groups, firstly as a test bed for the full theory and secondly as a class of new lattice theories possibly featuring an analogue diffeomorphism symmetry.

To make these notes accessible to readers outside the quantum gravity community we provide an introduction to some essential concepts in the loop quantum gravity, spin foam and group field theory approach and point out the many connections to lattice field theory and condensed matter systems.


\end{abstract}


\section{Introduction}

Spin foam models \cite{reisen, baezG,spinfoams} arose as one possible quantization method for gravity. These models can be seen in the tradition of lattice approaches to quantum gravity \cite{lollreview}, in which gravity is formulated as a statistical system on either a fixed or fluctuating lattice. Lattice methods are also ubiquitous in condensed matter and (Yang Mills) gauge theories. Indeed spin foam like structures appear in the strong coupling expansion of QCD or in the high temperature expansion for Ising gauge \cite{wegner} systems. 

In any lattice approach to gravity one has to discuss the significance of the (choice of) lattice for physical predictions. Although this point arises also for other lattice field theories, it carries extreme importance for general relativity. For matter fields on a lattice we can interpret the lattice itself as providing the background geometry and therefore background space time. In contrast, in general relativity, the geometry, and therefore space time itself is a dynamical variable, and has to be the outcome instead of one of the ingredients of the theory. This is one aspect  of  `background independence'  for quantum gravity \cite{rovellibook,thomasbook}.

One possibility to alleviate dependence of physical results on the choice of lattice, is to turn the lattice itself into a dynamical variable and to sum over (a certain class of) lattices. This is one motivation for the dynamical triangulation approach \cite{dyntriang}, causal dynamical triangulations \cite{causal} and group field theory \cite{bou,deppet}. See also \cite{knopf} for another possibility to introduce a dynamical lattice. The class of triangulations to be summed over can be restricted for instance in order to implement causality \cite{causal,dariojoe} or to symmetry--reduce models \cite{hexagon}.\footnote{In some cases the resulting models might then be reformulated as (tiling) models on a fixed lattice \cite{hexagon}.}

Another possibility is to ask for models which are per se lattice or discretization independent. Such models might arise as fixed points of a Wilsonian renormalization flow and represent so--called perfect discretizations \cite{perfectqcd,perfect,seb}. For gravity this concept is intertwined  with the appearance of discrete representation of diffeomorphism symmetry in the lattice models \cite{dittrichr,bahrdittrich1,perfect,hoehn,seb}. Such a discrete notion of diffeomorphism symmetry can arise in the form of vertex translations \cite{ruth4d}: Discrete geometries can be represented by a triangulation carrying geometric data, for instance in Regge calculus \cite{regge}  the lengths of the edges in the triangulation. Vertex translations are then symmetries that act (locally) on the vertices of the triangulations by changing the geometric data of the adjacent building blocks. If the symmetry is fully implemented into the model, then this action should not change the weights in the partition function. 

Such vertex translation symmetries are realized in $3D$ gravity, where general relativity is a topological theory \cite{freidel_l}. This means that  the theory has no local degrees of freedom, only a topology dependent finite number of global ones. The first order formalism of $3D$ gravity coincides with a $3D$ version of $BF$ theory \cite{bf_h}.  $BF$ theory is a gauge theory and can be formulated with a gauge group given by a Lie group. $3D$ gravity is obtained by choosing $SU(2)$. However, one can also choose finite groups \cite{witten_dijkgraaf,bais,mackaay}. In this case one obtains models used in quantum computing \cite{kitaev} as well as models describing topological phases of condensed matter systems, so called string net models \cite{wens}.  Moreover $BF$ systems can be seen as Yang--Mills like systems for a special choice of the coupling parameter (corresponding to zero temperature). Hence in $3D$ we have Yang Mills like theories with the usual (lattice) gauge symmetry as well as topological $BF$ theories with the additional translation symmetries.

$BF$ theory (in any dimension) is a topological theory because it has a large gauge group of symmetries, which reduces the physical degrees of freedom to a finite number.  It is characterized not only by the usual (lattice) gauge symmetry determined by the gauge group but also by so--called translational symmetries, based on the $3$--cells (i.e. cubes)  of the lattice and parametrized by Lie algbebra elements (if we work with Lie groups). For $3D$ gravity these symmetries can be interpreted as being associated to the vertices of the dual lattice (for instance given by a triangulation) and are hence termed vertex translations. However for $4D$ the symmetries are still associated to the $3$--cells of the lattice and hence to the edges of the dual lattice or triangulation. As in general there are more edges then vertices in a triangulation, we obtain much more symmetries then the vertex translations one might look for in $4D$ gravity. To obtain vertex translations of the dual lattice in $4D$ one would rather need a symmetry based on the $4$--cells of the lattice.

In $4D$ gravity is not a topological theory. There is however a formulation due to  Plebanski \cite{plebanski}  that starts with $BF$--theory and imposes so--called simplicity constraints on the variables. These break the symmetries of $BF$--theory down to a subgroup that can be interpreted as diffeomorphsim symmetry (in the continuum). This Plebanski formulation is also the one used in many spin foam models. Here one of the main problems is to implement the simplicity constraints \cite{barrett_crane,eprl,alexandrov,ryan2} into the $BF$ partition function. The status of the symmetries is quite unclear in these models, however there are indications \cite{bahrdittrich1} that in the discrete the reduction of the $BF$ translation symmetries does not leave a sufficiently large group to be interpretable as diffeomorphisms.  Nevertheless if these models are related to gravity, then there is a possibility, that diffeomorphism symmetry as the fundamental symmetry of general relativity, arises as a symmetry for large scales. In the above mentioned method of regaining symmetries by coarse graining, one can then expect that this diffeomorphism symmetry arises as vertex translation symmetry (on the vertices of the dual lattice).

We want to emphasize that hence in $4D$ spin foam models are candidates for a new class of lattice models in addition to Yang-Mills like systems and $BF$--theory. This new class would be characterized by a symmetry group in between those for Yang-Mills theory (usual lattice gauge symmetry) and $BF$--symmetry (usual lattice gauge symmetry and translation symmetry associated to $3$--cells). Here we would look for a symmetry given by the usual lattice gauge symmetry and translation symmetries associated to $4$--cells.

As we will show in this work the $4D$ spin foam models, which as gravity models are based on $SO(4),SO(3,1)$ or $SU(2)$, can be easily generalized to finite groups (or more generally tensor categories \cite{oecklbook}). Although the immediate interpretation as gravity models is lost one can nevertheless ask whether translation symmetries are realized and if not how these could be implemented. This question is much easier to answer for finite groups than in the full gravity case. One can therefore see these models as a test bed for the full theory. This also applies for renormalization and coarse graining techniques which need to be developed to access the large scale limit of spin foams. With finite group models it might be in particular possible to access the many--particle (that is many simplices or building blocks in the triangulation) and small spin (corresponding to small geometrical size of the building blocks) regime\footnote{The small spin regime arises as the spin is just a label for representation of the group and for finite groups there is only a finite number of irreducible unitary representations.}. This is in contrast to the few particle and large spin (semi--classical) regime
\cite{asymptotics} which is accessible so far.

In the emergent gravity approaches  \cite{emergent} one attempts to construct models, which do not necessarily start from gravitational or even geometrical variables, but nevertheless show features, typical of gravity. The models proposed here can also be seen as candidates for an emergent gravity scenario. A related proposal are the truncated Regge models  in \cite{z2gravity}, in which the continuous length variables of Regge calculus are replaced by values in $\Zl_q$, with $q=2$ for `Ising quantum gravity'.

~\\
Apart from providing a wealth of test models for quantum gravity researchers another intention of this work is to give an introduction to some of the spin foam concepts and ideas to researchers outside quantum gravity. Indeed spin foams arise as graphical tools in high temperature expansions of lattice theories, or more generally in the construction of dual models. We will review these ideas in section \ref{ising} where we will discuss Ising like systems and introduce spin nets as a graphical tool for the high temperature expansion. Next, in section \ref{gauge} we will discuss lattice gauge theories with (Abelian) finite groups. Here spin foams arise in the high temperature expansion. The zero temperature limit of these theories gives topological $BF$ theories which we discuss in more detail in section \ref{top}. In section \ref{nonA} we detail spin foam models with non--Abelian finite groups. Following a strategy from quantum gravity we will also discuss so--called constrained models which arise from $BF$ models by implementing simplicity constraints , here in the form of edge projectors, into the partition function. This will in general change the topological character of the $BF$ models to non--topological ones.

We will then discuss in section \ref{transfer} a canonical description for lattice gauge theories and show that for $BF$ theories the transfer operators are given by projectors onto the so called physical Hilbert space. These are known as stabilizer conditions in quantum computing and in the description of string nets. Here we will discuss the possibility to alter the projectors in order to break some subset of the translation symmetries of the $BF$ theories.

Next, in section \ref{gft}  we give an overview of group field theories for finite groups. Group field theories are then quantum field theories on these finite groups and generate spin foams as Feynman diagrams. Finally, in section \ref{nonlocal}  we discuss the possibility of non--local spinfoams. We will end with an outlook in section \ref{out}.

\section{Ising like models}\label{ising}

In this section we will review the construction of dual models \cite{wannier,savit} for Ising like models. Similar techniques will lead to the spin foam representation for (Ising) gauge theories. The main ingredient is to perform a Fourier expansion of the couplings, which can be understood as functions of group variables. Here the group in question is $\Zl_q$, with $q=2$ for the proper Ising models.

Consider any regular or random lattice\footnote{The models can be extended to oriented graphs. We will however not be interested in the most general situation but assume that the lattice or more generally cell complex is sufficiently nice in order to construct the dual lattices or complexes. Later on we will also need to be able to identify higher dimensional cells ($2-$cells and $3-$cells) in the lattice.} in which the edges are oriented (the latter is not necessary if we consider $\Zl_2$, that is Ising models).  On such a lattice one can for instance define an Ising model with spins $g_v$ associated to the vertices and nearest neighbour interaction by defining a partition function
\ba\label{za1}
Z&=&\sum_{g_v}  \prod_e \exp( \beta  g_{s(e)}g^{-1}_{t(e)})
\ea
where $g_v=\pm 1$ and group product is the normal product (or addition mod $2$ if $g=0,1$ is used) and $\beta$ is a coupling constant, proportional to the inverse temperature. The action $S=\beta \sum_e  g_{s(e)}g_{t(e)}$ describes the coupling between the spin $g_{s(e)}$ at the source or starting vertex and the spin $g_{t(e)}$ at the target or end vertex of every edge.

More generally we can assume that $g_v$ takes any values between $0, \ldots, q-1$ (giving Vector Potts models). The group product  corresponds now to addition modulo $q$. 
A further generalization of (\ref{za1}) is to allow for any edge weigth $w_e$ (even locally varying) as a function of the two group elements  $g_{s(e)}g^{-1}_{t(e)}$ associated to an edge $e$.
\ba\label{za2}
Z&=& \sum_{g_v} \prod_e w_e(g_{s(e)}g^{-1}_{t(e)}) \q .
\ea

Now the functions $w_e$ on the group can be expanded in a basis given by the irreducible representations of the group \cite{fulton-harris}. For the groups $\Zl_q$ this is just the usual discrete Fourier transform
\ba\label{za3}
w(g) &=& \sum_{k=0}^{q-1}   \tilde w(k) \, \chi_k(g)  \nn\\
\tilde w(k) &=& q^{-1} \sum_{g=0}^{q-1}  w(g)  \overline{\chi}_k(g)
\ea
where $\chi_k(g)=\exp(\frac{2\pi i}{q} k \cdot g)$ are the group characters. Characters for Abelian groups are multiplicative i.e. $\chi_k(g_1\cdot g_2)=\chi_k(g_1) \cdot \chi_k( g_2)$ and also $\chi_k(g^{-1})=\chi_k^{-1}(g)=\overline{\chi}_k(g)$.

Applying the character expansion to the weights in (\ref{za2}) we obtain
\ba\label{za4}
Z&=&\sum_{g_v} \prod_e \sum_{k} \tilde w_e(k) \chi_k(g) \nn\\
&=&\sum_{g_v} \sum_{k_e} \left(\prod_e  \tilde w_e(k) \right) \prod_e \chi_{k_e}(g) \nn\\
&=& \sum_{g_v} \sum_{k_e} \left(\prod_e  \tilde w_e(k_e) \right)  \prod_v   \prod_{e\supset v} \chi_{k_e}( g_v^{o(v,e)})
\ea
where $o(v,e)=1$ if $v$ is the source of $e$ and $-1$ if $v$ is the target of $e$. Here we just resorted the product such that in the last line the factor   $\prod_{e\supset v} \chi_{k_e}( g_v^{o(v,e)})$ involves all the appearences of $g_v$. We can now sum over $g_v$ which according to the Fourier inversion theorem gives a delta function involving the $k_e$ of the edges adjacent to $v$:
\ba\label{za5}
Z=q^{\sharp v} \sum_{k_e}  \left(\prod_e  \tilde w_e(k_e) \right)  \,\, \prod_v   \delta^{(q)} \!\big( \sum_{e\supset v} o(v,e) k_e\big)
\ea
where $\sharp v$ denotes the number of vertices and $ \delta^{(q)}(\cdot)$ is the $q-$periodic delta function.

We have now a representation of the partition function in which the group variables are replaced by variables taking values in the representation labels (that is the Pontrjagin dual of $G$). For $\Zl_q$ the Pontrjagin dual and the group itself are isomorphic, as both $g,k=0,\ldots,q-1$.  We will call a representation of the form (\ref{za5}) a spin foam\footnote{Note that in this particular example we are abusing language twice as  we do have neither `spin' nor a `foam'. The term spin arises in the full models from using the representation labels (spin quantum numbers) of $SU(2)$. The proper `foams' arise for lattice gauge theories as a higher dimensional generalization of the spin nets introduced in this section.} representation: there the sum over group elements is replaced by a sum over representation labels. Representations meeting at vertices (for gauge theories it will be edges) have to satisfy a certain condition, namely that the oriented sum of the representation labels has to vanish. This means that the tensor product of the representations meeting at an edge has to be trivial. For non--Abelian groups this condition is generalized to include some additional data: namely a projector from the tensor product of all representations meeting at a vertex into the trivial representation.

Note that the number of representation labels $k_e$ which are associated to the edges is in general different from the number of group variables $g_v$ attached to the vertices. But in addition we have the restriction that the sum of the representation labels meeting at a vertex has to be zero (mod $q$).  These are the Gauss constraints, demanding that the analogue of the electric field -- namely the assignment of representation labels to the edges, is divergence free.

To finalize the construction of dual models \cite{savit} one solves the Gau\ss~constraints and replaces the representation labels associated to the edges of the lattice by variables defined on the dual lattice. In two dimensions a divergence free vector field can be constructed from a scalar field ($v_1=\partial_2 \phi, v_2=-\partial_1 \phi$). On the lattice a similar construction leads to a dual model with variables associated to the vertices of the dual lattice.\footnote{Here we assume trivial cohomology of the lattice, otherwise one may find so--called topological modes, see \cite{rakowski} } Indeed for $\Zl_2$ one finds the Ising model again, just that the high (low) temperature regime of the dual model is mapped to the low (high)  temperature regime of the original model.

In three dimensions a divergence free vector field can be constructed from another vector field by taking its curl: $v_i=\epsilon_{ijk}\partial_j w_j$. However, adding a gradient $\partial_i \phi$ of a scalar field to the vector field $w_i$ will not change the values of the divergence free vector field $v_i$. This will be translated into a local gauge symmetry in the dual lattice. For the dual model one hence obtains as variables labels on the edges of the dual lattice  and the weights are functions of the holonomies around the plaquettes  (more generally faces, that is the two--dimensional cells) of the dual lattice, that is the couplings are between the variables associated to the edges around each face.

This results in a dual model which is a gauge model with variables associated to the edges of the dual lattice and with local gauge symmetries defined on the vertices of the dual lattice (as in the continuum this gauge symmetry is parametrized by a scalar field). We will discuss such theories in more detail in section \ref{gauge}.

In four dimension, with a similar argument as in three dimensions, the dual model is a `higher gauge theory' -- the variables are now associated to the plaquettes or faces of the dual lattice and the weights define couplings between the plaquettes bounding the three--dimensional cells. 

Finally in one dimension every vertex is adjacent to two edges, hence the Gauss constraints in (\ref{za5}) force the representation labels to be equal (we assume periodic boundary conditions). We can easily find
\ba\label{za6}
Z&=& q^{\sharp v}\sum_{k} \left(\prod_e  \tilde w_e(k) \right)  \nn\\
&=&q^{\sharp v} \sum_{k} \tilde w(k)^{\sharp e}
\ea
 where in the last line we assumed that the couplings are homogenous on the lattice, that is do not depend on the position or orientation of the edge.

The spin foam representation is adapted to the high temperature expansion \cite{hamer}: Indeed if we consider the Ising model with standard couplings as in (\ref{za1}) we obtain for the coefficients $\tilde w_0,\tilde w_1$ in the character expansion
\ba\label{za7}
\tilde w_0 = \cosh \beta  \q ,\q\q \tilde w_1 = \cosh \beta \tanh \beta  \q .
\ea
One can now expand the partition function (\ref{za5})  in powers of $\tilde w_1/\tilde w_0$, that is in the number of times the representation label $k_e=1$ appears.

Because of the Gau\ss~constraints in (\ref{za5}) the number of `excited' edges, that is carrying representation label $k_e=1$, entering each vertex has to be even, so for a cubical lattice the lowest order contribution is described by one plaquette $\sim (\tilde w_1/\tilde w_0)^4$.  This has to be weighted by a combinatorical factor, namely the number of ways one can embed the square into the lattice.

In general the terms in the perturbation series can be represented by closed polymers or graphs embedded into the lattice. The expansion in terms of these graphs can then be organized in different ways \cite{hamer}. Consider one particular contribution to the expansion, i.e. a configuration where a given number of edges carries non--trivial representation labels. In general such a contribution decomposes into connected components of edges with non--trivial representation labels, where we define two set of edges with $k_e\neq 0$ as disconnected if these sets do not share a vertex.

We will call such a connected component a (restricted) spin net, where restricted refers to the condition, that all edges have to carry a non--trivial representation. Because of the delta--functions in (\ref{za5}) such a spin net does not have open ends, i.e. all vertices have either valence two or higher. (For the Ising model only even valency is allowed.) Furthermore edges meeting at a two--valent vertex have to carry the same representation label. Note that this also holds for two--valent vertices, at which the edges `change direction'.  (Here we assume that the couplings $c_e$ or $w_e$ for the edges do not depend on the position or on the direction of the edges. Otherwise one has to equip the spin nets with more embedding data.) The representation labels can only change at so--called branching points where three or more edges meet. This allows to define so--called geometric (restricted) spin nets,\footnote{This definition is taken over from \cite{conrady1}, in which a similar concept was defined for spin foams, that is branched surfaces, which we will discuss in the next section. The following discussion is motivated by similar considerations for `spin foams' appearing in lattice  gauge theories \cite{zuber,reisen,baezG,conrady1}.} which are just specified by the connectivity of the spin net vertices (i.e. vertices of valence  higher than two with repect to the spin net) and the length of the edges of the spin net, that is the number of lattice edges that make up a spin net edge. The contribution of such a spin net to the partition function (or the free energy which would involve a sum over connected components) can be split into two parts, one is specific to the spin net itself, and from (\ref{za4}) given by
\ba\label{za8}
\prod_{e:k_{e}\neq 0} \tilde w_e(k_e) \q ,
\ea
the other contributions are combinatorical factors determined by the number of embeddings of the spin net into the given lattice. These factors could be summarized as measure (or entropic) factors, which carry the information of a given lattice, including its dimension. The spin net specific contribution factorizes over the edges of the spin net, for an edge with representation $k$ it is given by $\tilde w(k)^l$ where $l$ is the length of the spin net edge (i.e. the number of lattice edges that make up this spin net edge).

The combinatorical factors are determined by the `background' lattice and its structure. For quantum gravity models one would rather be interested in 'background independent' models, where such factors should not play a role. Here one can define an alternative partition sum over geometric spin nets, whose edges carry representation labels as well as another label, specifying the lengths of the edges. Again, in gravity the geometry of space time is a dynamical variable, i.e. should rather be encoded in the labels. This motivates to look for models, for which the spin net weights  are independent of the lengths of the spin net edges so that the weights depend only on the labels. These can be termed abstract spin nets \cite{conrady1}. (Note that with the definition we have taken here, the abstract spin nets will in general still remember some features of the lattice, for instance the valency of the spin net cannot be higher than the valency of the lattice.) For the Ising model such abstract spin nets arise in the limit $\tanh \beta =1$ for zero temperature. Indeed the high temperature expansion is an expansion in the sum of the length of the excited edges, whereas the abstract spin networks arise in a regime where this length does not play a role.

A similar structure arises in string net models \cite{wens}, which were designed to describe condensed phases of (scale free) branching strings. Indeed these string net models are closely related to (the canonical formulation of) topological field theories, such as $BF$--theories, which we are going to describe in section \ref{transfer}. In this canonical formulation spin networks will appear as one choice of basis for the Hilbert space of $BF$--theories. String net models assign amplitudes (corresponding to the so--called physical wave functions in $BF$--theory) to spin networks that satisfy certain (gauge) symmetry properties. These imply that these spin networks can be freely deformed on the lattice without changing their amplitude.

In the next section we will consider gauge systems, for which we will obtain a spin foam picture. Spin foams are similar to spin nets, just that the role of the edges is taking over by plaquettes or faces and the role of the vertices by edges or branchings. Also there one can define geometric and abstract spin foams \cite{conrady1}, which are branched surfaces \cite{zuber,reisen,baezG} whose elementary surfaces, or spin foam faces, carry a non--trivial representation label. (For non--Abelian groups the branchings or spin foam edges carry intertwiners between the representations associated to the surfaces meeting at the edges.) The amplitudes for the geometric spin foams will in general depend on the area of its elementary surfaces, i.e. the number of plaquettes making up the elementary surface in question, and the lengths of the branchings. This is for instance the case for Yang--Mills like theories, where geometric spin foams arise in the strong coupling or high temperature expansion.

The conditions for having abstract spin foams, that is amplitudes independent of the area of the surfaces, can be understood as requiring independence of the spin foam amplitude under trivial (face and edge) subdivision. This can be used to specify measure factors for the spin foam models \cite{reisen,conrady1,bahrk}.  Abstract spin foams will be also generated by group field theories discussed in section \ref{gft}, however the difference is that for the spin foams generated in this way trivial representation labels are still allowed. (Also there is not a restriction that these spin foams are embeddable into a given lattice.) See also \cite{matteo} for one possible connection between partition functions with restricted (i.e. trivial representations are not allowed) and unrestricted spin foams.


The discussion on geometric and abstract spin nets and spin foams assumes that the couplings do not depend on the position or direction of the edges or plaquettes in the lattice, otherwise one would need to introduce more decorations for the spin nets and foams. However one can show \cite{briegel} that for instance the $\Zl_2$ Ising gauge model in $4D$ with varying couplings is universal, i.e. can encode all other $\Zl_p$ lattice models. Here it would be interesting to develop the equivalent spin foam picture and to see how a spin foam with additional decorations could encode other spin foam models.

\section{Gauge theories}
\label{gauge}

The spin foam representation originated as a representation of partition functions for gauge theories \cite{reisen,baezG,rovellibook}. Such gauge theories can also be formulated with discrete groups \cite{wegner, kogut, bais} and for $\Zl_2$ give the Ising gauge models \cite{wegner}. A spin foam representation for Yang--Mills theories with continuous Lie groups has been presented in \cite{oecklpf}, see also \cite{conrady2},  and the results can be easily adapted to discrete groups. For abelian (discrete and continuous) groups these are again closely related to the construction of dual models \cite{savit} and the high temperature or strong coupling expansion \cite{zuber}.

Given a lattice (more precisely an orientable two--complex, in which we can uniquely specify the 2--cells, called faces or plaquettes) we associate group variables to the (oriented) edges of this lattice. We want to construct models with gauge symmetries at the vertices, where as usual the gauge action is given by $g_e \rightarrow g_{s(e)} g_e g_{t(e)}^{-1}$ where $g_v$ are gauge parameters associated to the vertices of the lattice (remember that $s(e)$ is the source vertex of $e$ and $t(e)$ the target vertex.). Gauge invariant quantities are given by (traces in some matrix representation of the group over) closed Wilson loops. Expressing the action as a function of these, the action will be gauge invariant, as can be easily checked.  The `smallest' Wilson loops are the loops made up of edges around a face $h_f=: \vec \prod g_e$ where here we take the oriented product, i.e. the orientation of the edges is the one induced by the orientation of the face (and $g_{e}=g_{e^{-1}}^{-1}$ where $e^{-1}$ is the edge with opposite orientation to $e$).

Given a unitary (finite dimensional) matrix representation of a group $g\mapsto U(g)$  a (Wilson--) action can be defined as
\ba
S&=& \alpha \sum_{f} \text{Re} \bigg(\text{tr} \big((U( h_f)\big)\bigg)  \q .
\ea
with $\alpha$ a coupling constant. For the groups $\Zl_q$ the irreducible unitary representations are one--dimensional and labelled by $k=0,\ldots, q-1$ so that the representations are given by $g \mapsto \exp( \frac{2\pi i}{q}\, k \cdot g)$.  (Here $g=0,\ldots q-1$ and the group product is given by addition.)

More generally we can assume the partition function to be of the form
\ba\label{y2b}
Z&=&\frac{1}{q^{\sharp e}}\sum_{g_e} \prod_f w_f(h_f)
\ea
where the weights $w_f$ are class functions, i.e. are invariant under conjugation $w_f(ghg^{-1})=w_f(h)$.

\subsection{Spin foam representation}\label{gaugesf}

The (exponential of the) action, and more generally the weights $w_f$  are class function of the plaquette variables $h_f$ and can hence be expanded in characters. For the Abelian groups $\Zl_q$ this again just amounts to the discrete Fourier transform. We can therefore consider the general partition function (where we include a measure normalization factor $1/q$ for every integration over the group $Z_q$).
\ba\label{y3}
Z&=&\sum_{g_e} \frac{1}{q^{\sharp e}} \prod_f  \sum_k \tilde w_{k} \,\,  \chi_k(h_f)  \q .
\ea
The derivation of the spin foam representation proceeds similar to the previous section \ref{ising}:
 \ba\label{y4}
Z&=&\sum_{g_e} \frac{1}{q^{\sharp e}}   \sum_{k_f}  \prod_f   \tilde w_{k_f}  \,\,  \chi_{k_f}(h_f) \nn\\
&=&  \frac{1}{q^{\sharp e}}    \sum_{k_f}  \left(  \prod_f \tilde w_{k_f} \right)  \prod_e \sum_{g_e}    \,\, \prod_{f\supset e} \chi_{k_f}(g_e^{o(f,e)}) \nn\\
&=&   \sum_{k_f}  \left(  \prod_f \tilde w_{k_f} \right)  \prod_e  \,  \delta^{(q)} \!\big( \sum_{f\supset e} o(f,e)k_f\big) \nn\\
&=&   \sum_{k_f}  \left(  \prod_f \tilde w_{k_f} \right)  \prod_v \prod_{e\supset v}  \delta^{(q)} \!\big( \sum_{f\supset e} o(f,e)k_f\big)
\ea
where $o(f,e)=1$ if the orientation of the edge coincides with the one induced by the (adjacent) face, and $-1$ otherwise. In the last step we just splitted the delta functions over every edge into two delta functions over every vertex, as in spin foams one usually works with vertex amplitudes (which here would be $ A_v=\prod_{e\supset v}  \delta^{(N)} \!\big( \sum_{f\supset e} o(f,e)k_f$). (These amplitudes and the splitting are more complicated for non--Abelian groups, see section \ref{nonA}.)

\subsection{Dual models}

Again the construction of the dual models \cite{savit} proceeds by solving for the delta functions associated to the edges in (\ref{y4}), that is by solving for the  Gau\ss~constraints. These enforce that the oriented sum of representation labels associated to the faces around one edge is vanishing. For the lowest dimensional case, namely two dimensions we obtain a 'trivial' dual theory. Since every edge is bounded by two faces, all the representation labels coincide $k_f \equiv k$ and we obtain (assuming the $\tilde w_{k_f}$ do not depend on the position and orientation of the plaquette, $\tilde w_{k_f}=\tilde w_k$)
\ba
Z= \sum_{k}  (\tilde w_{k})^{\sharp f}  \q .
\ea

For the three dimensional case we have seen that the models without any gauge symmetry are dual to a gauge model. Hence the Ising gauge model in $3D$ is dual to the Ising model in $3D$.  In $4D$ the gauge models map again to  gauge models. For nontrivial cohomology of the lattice topological modes in the dual model might appear \cite{rakowski}.



\subsection{High temperature expansion and spin foams}

As for the Ising models we can also discuss the high temperature expansion, that is for the Ising gauge model an expansion in powers of $\tilde w_1/\tilde w_0= \tanh \beta$.  The discussion is very similar to the one for the Ising models, the difference being that the perturbative contributions are now represented by closed surfaces and not closed polymers any more.

The `ground state' is the configuration $k_f=0$  
for all faces. The configurations with some $k_f  \neq 0$ have to satisfy the Gau\ss~constraints, which demands that the number of excited plaquettes (with $k_f\neq 0$) around every edge has to be even. This means in particular that the plaquettes with $k_f \neq 0 $ have to make up closed surfaces. For instance for the Ising gauge model on a hypercubical lattice one would obtain that the lowest order contribution would be represented by a ($3D$) cube giving an order $(\tilde w_1/\tilde w_0)^6$ for its six faces (times a combinatorical factor giving the numbers of embeddings of the cube into the lattice).

More generally the perturbative contributions are now described by closed and branched surfaces. (For the free energy one has only to consider connected components \cite{muenster}.) Here a proper branching would be a sequence of edges where at least three excited plaquettes meet. Excited plaquettes make up the elementary surfaces or spin foam faces. That is at the inner edges of these spin foam faces there are always only two excited plaquettes meeting. Again because of the Gau\ss~constraints, the representation labels of all the plaquettes in a given spin foam face have to agree. For homogeneous and isotropic couplings the spin foam amplitude (ignoring the combinatorical embedding factors) depends only on the number of plaquettes $a$ inside each spin foam face, the contribution of a spin foam face with label $k$ being $\sim (\tilde w_k/\tilde w_0)^a$, but not on how the spin foam is embedded into the lattice. (In general  for models  with non--Abelian groups the amplitudes might also depend on the length of the branching or spin foam edges.)  The definition of geometric and abstract spin foams \cite{conrady1} parallels the one for spin nets. Note that the condition for abstract spin foams invalidates the regime of the high temperature expansion, which is an expansion in the number of excited plaquettes of the underlying lattice.

In section \ref{top} we will discuss $BF$--theory, which is a topological model and which satisfies the conditions for an abstract spin foam, i.e. the amplitude does not depend on the areas $a$ of the spin foam faces. The reason is that $\tilde w_k/\tilde w_0 \equiv 1$ for the $BF$--models. For the Ising gauge model this is the case at zero temperature\footnote{Indeed in the limit of zero temperature the partition function for the Ising gauge models enforces all plaquette holonomies to be trivial. This coincides with the partition function of $BF$ theories discussed in section \ref{top}.}.

\subsection{Expansion around a topological sector}

If we start from the first line in (\ref{y3}) again and keep the summation over both group and representation labels we obtain
\ba\label{y5}
Z&=&\sum_{g_e} \frac{1}{q^{\sharp e}} \prod_f  \sum_k \tilde w_{k} \,\,  \chi_k(h_f)  \nn\\
&=& \frac{1}{q^{\sharp e}}  \sum_{g_e} \sum_{k_f}  \prod_f  \tilde w_{k_f} \exp(\frac{2\pi i}{q}  k_f \,\cdot\, h_f(g_e))  \nn\\
&=&\frac{1}{q^{\sharp e}}  \sum_{g_e} \sum_{k_f}  \prod_f  \exp(\frac{2\pi i}{q}  k_f \,\cdot\, h_f(g_e)  \,\, + \,\, \ln \tilde w_{k_f})  \q .
\ea
This corresponds to a first order representation of Yang--Mills theory where $g_e$ are the connection variables and $k_f$ represent the dual (electric field or flux) variables. The model where all $\tilde w_{k_f}=1$ is a topological $BF$--model, whose partition function can be solved exactly. Hence Yang--Mills theory can be understood as a deformed $BF$--theory \cite{martinelli,conrady2} -- the deformation appears here as the $\ln(\tilde w_{k_f})$ term (in the continuum it is $B\wedge \star B$).

The expansion of models with propagating degrees of freedom, for instance $4D$ gravity and Yang--Mills, around topological theories has been discussed for instance in \cite{artem, conrady1}

 For the Ising case we can obtain an expansion around the $BF$--partition function by introducing the expansion parameter $\alpha=\tanh\beta-1$ which is small for low temperature:
\ba\label{y6}
Z &=& (\cosh \beta)^{\sharp f} \sum_{k_f}
\prod_e  \,  \delta^{(q)} \!\big( \sum_{f\supset e} o(f,e)k_f\big)  \left(  \prod_f (\delta(k_f,0)+ \delta(k_f,1)(1+\alpha)\,) \right)  \nn\\
&=& (\cosh \beta)^{\sharp f} \sum_{k_f}
\prod_e  \,  \delta^{(q)} \!\big( \sum_{f\supset e} o(f,e)k_f\big)  \times   \nn\\
&&\q\q\bigg( 1+ \alpha \sum_{f_1} \delta(k_{f_1},1)  + \alpha^2 \sum_{f_1< f_2} \delta(k_{f_1},1) \delta(k_{f_2},1) +\ldots  \nn\\
&&\q \q\q\q\q\q\q\q\q\q\q \ldots+ \alpha^{\sharp f} \sum_{{f_1}<\ldots < {f_{\sharp f}}} \delta(k_{f_1},1)\cdots   \delta(k_{f_{\sharp f}},1) \bigg) \
 \q .
\ea
That is for the coefficient of $\alpha^1$ we have to count for each configuration $\{k_f\}_f$ the number of excited faces. For the coefficient of $\alpha^2$ we sum over ordered pairs of faces and in this way count for each configuration the number of ordered pairs of excited faces. The next is a sum over ordered triples of faces and so on. The last coefficient for $\alpha^{\sharp f}$ is one for the configuration $k_f\equiv 1$ and zero for all other configurations. 
The terms in this expansion can be seen as expectation values of observables in a $BF$--model, where the observable is the number of ordered $n$--tuples of excited plaquettes.  




\section{Topological models} \label{top}

In the last section we have seen that the Yang--Mills type theories can be understood as a deformation of $BF$--theories, which we will here discuss in more detail. $F$ stands for the curvature of a connection, usually taking values in the Lie group, and $B$ is a Lie algebra valued $(D-2)$--form. Hence in constructing these models for discrete groups the question arises with what to replace these variables $B$. We have seen that the field $B$ corresponds to the representation labels. Examples for topological models with discrete groups are discussed for instance in \cite{witten_dijkgraaf,birmingham,mackaay}. See also \cite{asano-higushi} for the relation between Chern-Simons like theories and $BF$-like theories for $\Zl_q$ in $3D$. In this section we will restrict to Abelian groups, in particular $\Zl_q$. The models for non--Abelian groups will be presented in section \ref{nonA}.

$BF$--theory is a topological field theory in which the equations of motion demand that the (local) curvature vanishes. We will start from the partition function encoding this requirement. In addition to the usual gauge theory symmetry the model enjoys an additional gauge symmetry involving the representation labels, which we will discuss later on.
To define the model assume that we have given a (oriented) two--complex (or for simplicity a lattice) with oriented edges $e$ which bound faces (or plaquettes) $f$. Associate to the edges group elements $g_e$. (With $g_{e}=g^{-1}_{e^{-1}}$, where $e^{-1}$ is the edge with reversed orientation to $e$.) Then we start from the following definition for a partition function $Z$ (see for instance \cite{baezG,baez_BF})
\ba\label{z1}
Z&=& \sum_{g_e}\frac{1}{q^{\sharp e}} \prod_f \delta_G (h_f)
\ea
where again $h_f=\vec\prod_{e \subset f}g_e$ means the oriented product of edges around a face.
Here we normalized the integration measure over the group to one $\sum_{g} \frac{1}{q} 1=1$ where $q$ is the cardinality of the group. $\delta_g$ is the delta function with respect to this measure, i.e.  $\sum_g \frac{1}{q}  \delta_G( g) f(g)=f(\text{id})$.
Note that this is just a special case of the gauge theory partition functions (\ref{y2b}), by setting $w_f=\delta_G$ there.

 The partition function can be easily evaluated to
\ba
Z=\sharp\left[\mbox{configurations $\{g_e\}_e$, such that}\q   \vec\prod_{e \subset f} \,\, g_e= \text{id} \q \forall \q f \right] \,\times\, \frac{q^{\sharp f}}{q^{\sharp e}}
\ea
where the second factor arises due to the normalization factors for the group integrations over $g_e$ and for the delta functions for each face.

For  the (Abelian) cyclic groups $Z_q$ we obtain by using the expansion $\delta_G=\sum_k \chi_k$
\ba\label{z4}
Z&=&q^{-\sharp e} \sum_{g_e} \prod_f \delta_G (\vec\prod_{e \subset f} \,\, g_e) \nn\\
&=&  q^{-\sharp e} \sum_{g_e} \prod_f  \sum_{k} \vec\prod_{e \subset f} \chi_k(g_e) \nn\\
&=& q^{-\sharp e}  \sum_{g_e}  \sum_{k_f}   \exp\left( \frac{2\pi i}{q}\, \sum_f k_f \cdot  \sum_{e\subset f}{o(f,e)} g_e\right) \q ,
\ea
where $o(f,e)=1$ if the orientation of $e$ agrees with the one induced from $f$ and $-1$ otherwise. The term in the exponential can be interpreted as a $BF$--action, with the $B$--field given by the assignment $f \mapsto k_f$ of representation labels to faces and the curvature $F$ by the holonomies $ h_f=\vec\prod_{e \subset f}g_e$. (In the exponential we used the representation of $\Zl_q$ as additive group.)


The spin foam representation can be derived in the same way as for the gauge theories in section \ref{gaugesf}:
\ba\label{z5}
Z&=&\frac{1}{q^{\sharp e}}  \sum_{k_f}  \prod_e \,\,  \sum_{g_e}  \prod_{f \supset e} \chi_{k_f}(g_e^{o(f,e)}) \nn\\
&=& \frac{1}{q^{\sharp e}}  \sum_{k_f} \prod_e \,\,   q \, \delta^{(q)}\!(\,\sum_{f \supset e} o(f,e) k_f )
   \nn\\
&=& \sum_{k_f} \prod_v\,\,  \prod_{e \supset v}  \delta^{(q)}\!\big(\,\sum_{f \supset e} o(f,e) k_f \big) \q .
\ea
This gives $Z$ as
\ba\label{z6}
Z&=&\sharp\big[\mbox{configurations $\{k_f\}_f$, such that}\q  \sum_{f \supset e} o(f,e) k_f  = 0 \;\; \text{mod}\;q\q  \forall \q e \big]
\ea
that is as the number of assignements $\{k_f\}_f$ satisfying the Gau\ss~constraint for every edge. In the next section we will discuss the local symmetries of the $BF$ partition function. We will find that on the $\{k_f\}_f$ there is a translation symmetry acting which is based on the 3--cells of the lattice. Hence $Z$ reflects the orbit volume of this translational gauge symmetry. This leads to divergencies for the $BF$ partition functions for (continuous) Lie groups \cite{mb}.\footnote{However not all divergencies are due to this gauge symmetry \cite{mb}.}






\subsection{Symmetries of the partition function}\label{symm}

Here we will discuss the gauge symmetries of the partition function (\ref{z1}). As it is of the form of a gauge theory (\ref{y2b}) the partition function is invariant under the usual
gauge transformations: $g_e \rightarrow g_{s(e)}g_{e}g_{t(e)}^{-1}$, where $g_v \in G$ are gauge parameters associated to the vertices. This is obvious in the representation (\ref{z1}) for the partition function.

There is a further symmetry, the so--called translation symmetry, which is easiest to see in the spin foam representation.  This symmetry is based on the $3-$cells $c$ of the lattice, i.e. the cubes for a hypercubical lattice.\footnote{In $2D$ this symmetry degenerates into a global symmetry: the Gau\ss~constraints force all $k_f$ to be equal, $k_f\equiv k$, however the contribution of every representation label $k$ to the partition function is the same.} Consider a field $c \mapsto k_c$ of gauge parameters associated to the cubes and define the gauge transformations
\ba\label{z7}
k'_f=k_f+\sum_{c \supset f} o(c,f) k_c  \q \text{mod}\;q  \q
\ea
where $o(c,f)=1$ if the orientation of $f$ agrees with the one induced by $c$ and $-1$ otherwise. Then if $\{k_f\}_f$ is a configuration which satisfies all the Gau\ss~constraints, so is $\{k'_f\}_f$ and vice versa. The reason is that if an edge $e$ is in the boundary of a $3-$cell $c$ then there are exactly two faces $f,f'$ which are both in the boundary of $c$ and adjacent to $e$.  The orientation factors are such, that the contributions of $k_c$ from the two faces $f,f'$ are of opposite sign in the Gau\ss~constraint for $e$. Hence its value does not change, and therefore the contribution of the two configurations $\{k_f\}_f, \{k'_f\}_f$ to the partition function is equal.

This is the principle, that `the boundary of a boundary is zero' $\partial \circ \partial=0$, which underlines the Bianchi identity for the curvature. With the help of this Bianchi identity one can also explain the translation symmetry, see  \cite{freidel_l,dittrichr}. For $3D$ gravity this translation symmetry corresponds to the diffeomorphism symmetry of the theory. The gauge field $k_c$ corresponds in the dual lattice to a field associated to the dual vertices and one can interpret the gauge transformation as a translation\footnote{the gauge parameters are then lie algebra valued fields and for the lie algebra $su(2)$ indeed correspond to translations} of these dual vertices (which in the gravity models are the vertices in a triangulation of space time).

In $4D$ the $3-$cells are dual to edges in the dual lattice and the translation symmetries rather translate these instead of the dual vertices. However for gravity--like models one would need to break down the translation symmetries from the $BF$--case, where it is based on the dual edges, to symmetries based on the dual vertices. (Correspondingly in $4D$ diffeomorphism symmetry of the action leads to Noether charges encoding the contracted Bianchi identities and not the Bianchi identities itself.) In the case of gravity one strategy is to impose so--called simplicity constraints \cite{plebanski,barrett_crane,eprl,ryan2} into the partition function. For the models with finite groups the question arises, if there is a class of $4D$ gauge models with (translation like) symmetries based on the dual vertices. Such models would be non--topological and feature propagating degrees of freedom, as the symmetry group based on dual vertices is much smaller than for $BF$, where it is based on dual edges. See also the discussion in sections \ref{nonA} and \ref{transfer}.




\section{Gauge theories for non--Abelian groups} \label{nonA}

In the following, we will consider how to generalize the construction performed so far for finite, but non-Abelian groups $G$. Details about representations can be found in \cite{fulton-harris}.

Again, we denote by $\rho$ the irreducible, unitary representations of $G$ on $\mathbb{C}^n\simeq V_\rho$ where $n= \text{dim}\,\rho$. Note that for non-Abelian groups $n$ can be larger than $1$. Every function $f:G\to\mathbb{C}$ can be decomposed into matrix elements of representations, i.e.
\begin{eqnarray}\label{Gl:CharacterExpansion}
f(g)\;=\;\sum_\rho \sqrt{\dim\rho}\,\tilde f_{\rho a b}\,\rho(g)_{ab}
\end{eqnarray}

\noindent where the sum ranges over all (equivalence classes) of irreducible representations of $G$. The $\tilde f_\rho$ are given by
\begin{eqnarray}
\tilde f_{\rho,a,b}\;=\;\frac{1}{|G|}\sum_g\sqrt{\dim\rho}\,f(g)\,\rho(g)_{ab}
\end{eqnarray}
where $|G|$ denotes the number of elements in the group $G$. Introducing an inner product between functions $f_1,f_2:G\to\mathbb{C}$ by
\begin{eqnarray}
\langle f_1\,|\,f_2\rangle\;=\;\frac{1}{|G|}\sum_{g\in G}\overline{f_1(g)}\,f_2(g)\;=\;\sum_{\rho,a,b}\overline{(\tilde {f}_1)_{\rho,a,b}}(\tilde{f}_2)_{\rho,a,b}
\end{eqnarray}

\noindent then the matrix elements of $\rho$ are orthogonal, i.e.
\begin{eqnarray}
\langle \rho_{ij}\,|\,\tilde\rho_{kl}\rangle\;=\;\frac{\delta_{\rho,\tilde\rho}\,\delta_{ik}\,\delta_{jl}}{\dim \rho} \q .
\end{eqnarray}

\noindent Furthermore, we denote by
\begin{eqnarray}
\chi_\rho(g)\;=\;\text{tr}(\rho(g))
\end{eqnarray}

\noindent the character of $\rho$, then the $\delta$-function on the group $G$ is given by
\begin{eqnarray}
\delta_G(g)\;=\;\sum_{\rho}\dim\rho\;\chi_\rho(g)
\end{eqnarray}

\noindent which satisfies
\begin{eqnarray}
\frac{1}{|G|}\sum_g\delta_G(g)\,f(g)\;=\;f(1)
\end{eqnarray}

\subsection{$G$--gauge theory on a two-complex}

\begin{figure}[hbt!]
\begin{center}
    \psfrag{f}{$f$}
    \psfrag{e1}{$e_1$}
    \psfrag{e2}{$e_2$}
    \psfrag{e3}{$e_3$}
    \psfrag{e4}{$e_4$}
    \psfrag{e5}{$e_5$}
    \includegraphics[scale=0.5]{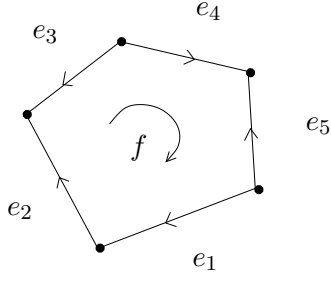}
    \caption{A face $f$, bordered by edges $e_1,\ldots, e_5$. The curvature is given by $h_f=g_{e_5}^{-1}g_{e_4}g_{e_3}^{-1}g_{e_2}g_{e_1}$ (note the ordering).}\label{Fig:Curvature}
\end{center}
\end{figure}

Consider an oriented two-complex\footnote{More specifically, we consider only complexes which are such that the glueing maps used to successively build up $\kappa$ are injective. This in particular means that the boundary of each face contains each edge at most once. With certain choices for amplitudes, this condition can be relaxed slightly, see e.g. \cite{bahrk}} $\kappa$. Denote the set of edges $E$ and the set of faces $F$, then by a \emph{connection} we mean an assignment $E\to G$ of group elements $g_e$ to edges $e\in E$. For every face $f\in F$, we denote the \emph{curvature} of this connection by
\begin{eqnarray}
h_f\;:=\;\overrightarrow{\prod_{e\in\partial f}}g_e^{o(f,e)}
\end{eqnarray}
\noindent the ordered product of group elements of edges in the boundary of $f$, where we take $o(f,e)=\pm1$ depending on the relative orientation of $e$ and $f$. In the non-Abelian case the ordering of the group elements around a face/plaquette actually matters, and we choose here the convention as depicted in  figure \ref{Fig:Curvature}.


As before we define a gauge invariant partition function by choosing a collection of weight-functions $w_f:G\to \mathbb{C}$ invariant under conjugation. These encode the action (and path integral measure) of the system. The partition function is defined as
\begin{eqnarray}\label{ben1}
Z\;=\;\frac{1}{|G|^{\sharp e}}\sum_{g_e}\,\prod_{f\in F}\,w_f(h_f)
\end{eqnarray}

\noindent where $|G|$ is the number of elements in $G$. Since $w_f$ can be expanded into characters via (\ref{Gl:CharacterExpansion}), the path integral can be written as
\begin{eqnarray}\label{Gl:StateSum2}
Z\;=\;\frac{1}{|G|^{\sharp e}}\sum_{g_e}\sum_{\rho_f} \prod_f(\tilde w_f)_\rho\;\rho_f(g_{e_1}^{\pm1})_{i_1i_2}\rho_f(g_{e_2}^{\pm1})_{i_2i_3}\cdots \rho_f(g_{e_n}^{\pm1})_{i_ni_1}
\end{eqnarray}

\noindent In the sum there is for each edge $e$ a representation $\rho_f(g_e)$ appearing for every face $f$ adjacent to e, $f\supset e$. The corresponding sum results in (assuming the orientations of all $f$ agree with that of $e$ for the moment, in order not to overburden the notation)
\begin{eqnarray}\label{Gl:DefinitionOfProjector}
(P_e)_{i_1i_2\cdots i_n;\;j_1j_2\cdots j_n}\;:=\;\frac{1}{|G|}\sum_{g\in G}\rho_{f_1}(g)_{i_1j_1}\rho_{f_2}(g)_{i_2j_2}\,\cdots\,\rho_{f_n}(g)_{i_nj_n}
\end{eqnarray}

\noindent It is not hard to see that the operator $P$, called the \emph{Haar-intertwiner},  given by
\begin{eqnarray}
(P_e\psi)_{i_1i_2\cdots i_n}\;:=\;(P_e)_{i_1i_2\cdots i_n;\;j_1j_2\cdots j_n}\psi_{j_1j_2\cdots j_n} \q .
\end{eqnarray}

\noindent which maps the edge-Hilbert space
\begin{eqnarray}\label{Gl:DefinitionEdgeHilbertSpace}
\mathcal{H}_e\;:=\;V_{\rho_1}\otimes V_{\rho_2}\otimes\ldots\otimes V_{\rho_n}
\end{eqnarray}

\noindent to itself, is actually an orthogonal projector on the gauge-invariant subspace of $\mathcal{H}_e$. \footnote{This can be easily shown by proving $P_e=P_e^*$, resulting from the unitarity of all representations, and $(P_e)^2=P_e$, which uses the translation-invariance and normalization of the Haar measure on $G$.} Note that, while in the Abelian case one has to sum over all representations over faces such that at all edges the (oriented) sum adds up to zero (as in (\ref{y4})), in the non-Abelian generalization one has to sum over all representations such that at each edge $e$ the representation of the faces $f\supset e$ meeting at $e$ have to contain, in their tensor product, the trivial representation. Also, one obtains a nontrivial tensor $P_e$ for each edge, which in the case of Abelian gauge groups is just the $\mathbb{C}$-number $\delta_{\pm k_1\pm k_2\pm\cdots\pm k_n, 0}$. Since the representations for non-Abelian groups can be more than $1$-dimensional, in general the tensor $P_e$ has indices which are contracted at each vertex $v$ in the two-complex $\kappa$.

Choosing an orthonormal base $\iota_e^{(k)}$, $k=1,\ldots,m$ for the invariant subspace of $\mathcal{H}_e$, we get
\begin{eqnarray}\label{Gl:Decomposition}
P_e\;=\;\sum_{k=1}^m\;|\iota_e^{(k)}\rangle\langle\iota_e^{(k)}|
\end{eqnarray}

\noindent 
In case the orientations of $e$ and $f$ do not agree for some $e$, then $g_e$ is essentially replaced by $g_e^{-1}$ in (\ref{Gl:DefinitionOfProjector}), which leads to the appearance of the dual\footnote{Defined by $\rho^*(h)_{ab}=\rho(h^{-1})_{ba}$.} representation in the tensor product (\ref{Gl:DefinitionEdgeHilbertSpace}) of the $\mathcal{H}_e$. In this case $\iota_e^{(k)}$ labels a basis of intertwiner maps between the tensor product of all representations associated to faces $f$ with $o(f,e)=-1$ and the tensor product of all representations associated to the faces with $o(f,e)=1$.

\begin{figure}[hbt!]
\begin{center}
    \psfrag{v}{$v$}
    \psfrag{e1}{$e_1$}
    \psfrag{e2}{$e_2$}
    \psfrag{f}{$f$}
    \includegraphics[scale=0.55]{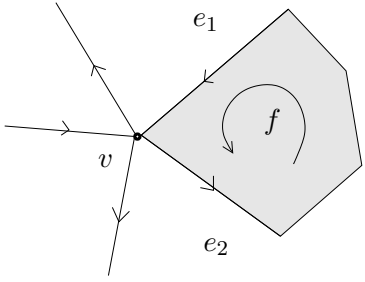}
    \caption{The relative orientations of $e_1$ and $f$, and $e_2$ and $f$ agree, so in both the edge Hilbert spaces $\mathcal{H}_{e_1}$ and $\mathcal{H}_{e_2}$ the factor $V_{\rho_f}$. Therefore $|\iota_{e_1}\rangle$ and $\langle\iota_{e_2}|$ have opposite indices belonging to $V_{\rho_f}$, which therefore can be contracted. }\label{Fig:Vertex}
\end{center}
\end{figure}

In (\ref{Gl:Decomposition}), one regards $|\iota_e\rangle$ and $\langle\iota_e|$ as attached to, respectively,  the endpoint and the beginning of $e$. For each vertex $v$ in $\kappa$, the associated $|\iota_e\rangle$ and $\langle\iota_e|$ can be contracted in a canonical way: For every face $f$ which touches $v$, there are exactly tho edges $e_1$, $e_2$ in the boundary of $f$ that meet at $v$. The definitions above are exactly such that, if one chooses a basis in each $V_\rho$ and the dual basis in each $V_\rho^*$, in $\iota_{e_1}$ and $\iota_{e_2}$ the two indices associated to $f$ are in opposite position, so the can be contracted. See figure \ref{Fig:Vertex} for an example.

Therefore, contracting all the appropriate $\iota_e$ at one vertex leaves one with the vertex-amplitude $\mathcal{A}_v(\rho_f,\iota_e)$, which depends on the representations $\rho_f$ and intertwiners $\iota_e$ associated to the faces and edges that meet at $v$ (and the orientations of these). The vertex amplitude can be computed by evaluating the so-called \emph{neighbouring spin network function}, living on graph which results from a dimensional reduction of the neighbourhood of $v$:  Construct a spin network, where there is a vertex for each edge $e$ touching $v$, and a line between any two vertices for each face $f$ between two edges. Assigning the $\rho_f$ to the lines of the spin network, and the intertwiners $|\iota_e\rangle$ or $\langle\iota_e|$, to the vertices, depending on whether the corresponding edge $e$ is incoming or outgoing of $v$, results in a spin network, whose evaluation gives $\mathcal{A}_v$.

With this, the state sum can, using (\ref{Gl:StateSum2}) and (\ref{Gl:DefinitionOfProjector}), be written in terms of vertex amplitudes via
\begin{eqnarray}\label{Gl:StateSum}
Z\;=\;\sum_{\rho_f\,\iota_e}\;\prod_{f}(\tilde w_f)_\rho\;\prod_v\mathcal{A}_v(\rho_f,\iota_e)
\end{eqnarray}

\subsection{Examples}

The first example we consider is $G$-$BF$ theory, which corresponds to an integral over only flat connections, i.e. the choice $w_f(h)=\delta_G(h)$. Since the $\delta$-function can be decomposed into characters with $(\tilde w_f)_\rho=\dim\rho_f$, we get
\begin{eqnarray}
\mathcal{Z}\;=\;\sum_{\rho_f\,\iota_e}\;\prod_{f}\dim\rho_f\;\prod_v\mathcal{A}_v \q .
\end{eqnarray}
\noindent If $\kappa$ is the two-complex dual to a triangulation of a $3$--dimensional manifold, the vertex amplitude is essentially the analogue of the $6j$-symbol for $G$\footnote{Not that there is some ambiguity in the literature about what is actually called the $6j$-symbol, which is a question of the correct normalization. We mean the normalized $6j$-symbol here, since we chose the intertwiners to be normalized in the first place. The normalization, which usually results in nontrivial edge-amplitudes, can be absorbed into the vertex amplitude. Also there might, according to convention, some sign factors be assigned to the edges $e$, which may not be absorbed into the vertex amplitudes $\mathcal{A}_v$.}.

\noindent The second example one usually considers is Yang-Mills theory. The Wilson action can be specified, similar to the Abelian case, by choosing a unitary representation $\tilde \rho$, so that
\begin{eqnarray}\label{Gl:WilsonianYangMillsAction}
S_{YM}(h)\;=\;\frac{\alpha}{2} \;\big(\chi_{\tilde\rho}(h)+\chi_{\tilde\rho}(h^{-1})\big)\;=\;\alpha \;\text{Re}\big(\chi_{\tilde\rho}(h)\big)
\end{eqnarray}

\noindent where $\alpha$ is the coupling constant. The weights are then given by $w_f(h)=\exp(-S_{YM}(h))$.

\subsection{Constrained models} \label{constra}

There is a generalization of the state--sum models (\ref{Gl:StateSum}), coming from the desire to obtain non--topological generalizations of $BF$-theory. It amounts to choosing the intertwiners $\iota_e$ not to span all of the invariant subspace, but only a proper subspace $V_e\subset \text{Inv}(\mathcal{H}_e)$. This originates in the attempt to define a state-sum model for general relativity, which can be written as a constrained $BF$-theory. The subspace $V_e$ is viewed as the space of intertwiners satisfying the so-called simplicity constraints, see e.g. \cite{plebanski, dittrichryan, ryan2}. Examples for such models are the Barrett-Crane model \cite{barrett_crane} and the EPRL  and FK  models \cite{eprl}. These models can also be written in the form of a path integral over connections \cite{Freidel:1998pt, oecklpf,pfeifferc}, but we shall not concern ourselves with this here.

In the following, we will introduce a class  of models which can be seen as the generalization of the Barrett-Crane models to finite groups. Given a finite group $H$, the model is a state-sum model for the group $G=H\times H$. Irreducible representations of $G$ are then pairs of irreducible representations $(\rho_f^+,\rho_f^-)$ of $H$. In the edge Hilbert-space $\mathcal{H}_e$ there is a specific element $\iota_{BC}$ of the space of gauge-invariant elements, called \emph{Barrett-Crane-intertwiner}. It is only nonzero iff the representations $\rho_f^+$ and $\rho_f^-$ are dual to each other. In that case, for an edge $e$ with attached faces $f_k$, consider the Hilbert space
\begin{eqnarray}
\tilde{\mathcal{H}}_e\;=\;V_{\rho_{f_1}^+}\otimes\ldots\otimes V_{\rho_{f_k}^+}
\end{eqnarray}

\noindent Again, we have assumed that the orientations of all faces $f_k$ and $e$ agree, in order not to overburden the notation, otherwise replace $V_{\rho_f^+}$ by its dual, or equivalently, exchange $\rho_f^\pm$. If we consider the projector $\tilde P_e:\tilde{\mathcal{H}}_e\to \tilde{\mathcal{H}}_e$ onto the subspace of elements invariant under the action of $H$, then $\tilde P_e$ can be seen as an element of $\tilde{\mathcal{H}}_e\otimes\tilde{\mathcal{H}}_e^*$, which is naturally isomorphic to $\mathcal{H}_e$, because $\rho_f^\pm$ are dual to each other. The corresponding element $\iota_{BC}$ in $\mathcal{H}_e$ is invariant under $H\times H$, as one can readily see, and it is our distinguished element, onto which $P_e$ is projecting.

Since the projection space for each edge is (at most) $1$-dimensional, the vertex amplitude for the BC-model does not depend on intertwiners, but only on the representations $\rho_f^+=(\rho_f)^*$ associated to the faces. In fact, since the representations are unitary, every representation is dual to itself, so we effectively have only one representation $\rho_f$ attached to each face. Using diagrammatical calculus \cite{birdtracks}, it is not hard to show that the vertex amplitude for the BC model is given by a sum over squares of the $BF$-theory model on the same vertex, i.e.
\begin{eqnarray}
\mathcal{A}_v^{(BC)}(\rho_f^\pm,\iota_{BC})\;=\;\sum_{\iota_e}\big|\mathcal{A}^{(BF)}_v(\rho_f,\iota_e)\big|^2
\end{eqnarray}

\noindent where the sum ranges over an orthonormal basis $\iota_e$ of $S_3$-intertwiners in $\tilde{\mathcal{H}}_e$ for every edge $e$ at the vertex $v$.

Let us shortly discuss the case of $H=S_3$, the group of permutations in three elements. For $S_3$, which is generated by $(12)$, the permutation of the first two elements, and $(123)$, which is the cyclic permutation, subject to the relation $(12)^2=(123)^3=1$, the representations theory is well-known \cite{fulton-harris}: There are three irreducible representations, called $[1]$, $[-1]$ and $[2]$. The first is the trivial representation, the second one maps a permutation $\sigma$ to its sign $(-1)^\sigma$. The third one is two-dimensional, and
\begin{eqnarray}
\rho((12))\;=\;\left(\begin{array}{cc}1&0\\0&-1\end{array}\right),\qquad\qquad
\rho((123))\;=\;\left(\begin{array}{cc}\cos2\pi/3&-\sin2\pi/3\\\sin2\pi/3&\cos2\pi/3\end{array}\right)
\end{eqnarray}

\noindent The nontrivial tensor products of the representations decompose as
\begin{eqnarray}
[-1]\otimes[-1]\;&=&\;[1],\qquad[-1]\otimes[2]\;=\;[2]\\[5pt]\nonumber
[2]\otimes[2]\;&=&\;[1]\oplus[-1]\oplus[2]
\end{eqnarray}

\noindent Therefore the intertwiner space in $\mathcal{H}_e$ is e.g. three-dimensional, when there are four faces attached to $e$ carrying the representation $[2]$. Hence, for $H=S_3$, the BC-model is an example for a constrained version of a $H\times H$-state-sum model, as defined in the last section, because $P_e$ projects onto a proper subspace of the invariant subspace of $\mathcal{H}_e$.\\

This class of models is an example for an abstract spin foam \cite{conrady1}, since its amplitudes only depend on combinatorial data, and the topology of the two-complex. In particular, it leads to a model which is invariant under refinement of the two-complex $\kappa$ by trivial subdivisions of edges or faces. Hence these models provide potential examples for models which are background-independent, but not topological (as is the case in the full theory).\\

Note that there is a wealth of different models, which come from different choices of nontrivial subspaces of $\mathcal{H}_e$, onto which $P_e$ in projecting. In particular, one could consider EPRL-like models where $G=S_4\times S_4$ or $G=A_4 \times A_4$ and consider the subspace of intertwiners for $\rho_f^+=\rho_f^-$, which are in the image of a boosting map $b:V_{\rho_f}\to V_{\rho_f}^+\otimes V_{\rho_f}^-$, given by the fusion coefficients (see e.g. \cite{eprl} for $H=SU(2)$). Here $S_4$ ($A_4$) is the  group of (even) permutation  of four elements. Another, possibly more geometric way would be to consider embeddings $S_4\to S_5$ and thus construct proper subspaces of $S_5$-intertwiners.\footnote{We are thankful to Frank Hellmann for this insight.} Such models can be considered as truncations of the EPRL models as $S_4$ ($A_4$) is the (chiral) symmetry group of the tetrahedron and a subgroup of the rotation group. Here it will be interesting to investigate the relation to the full models in more detail.


The models can serve to test many proposals for the full theory, for instance how the different choices of edge projectors $P_e$ determine the physical degrees of freedom or particle content of the given theory. This is related to the problem of implementing the simplicity constraints into spin foam models.
We are planning to investigate the features of these models, in particular its symmetries \cite{ta2} and behaviour under  coarse graining, in future work.

\section{Hilbert space, transfer operators and constraints}\label{transfer}

In this section we are going to derive the transfer operators  \cite{kogut_susskind, smit} for the Yang--Mills like gauge theories with partition function (\ref{y2b},\ref{ben1}) with a finite group $G$ with cardinality $|G|$. Here we will also allow for non--Abelian groups. The first part of the discussion will be similar to \cite{smit} for Yang-Mills theory, however we will also include $BF$--theory as a special case. From the transfer operators one can obtain in a limiting procedure the Hamiltonian operators. For $BF$--theory this limiting procedure will not be necessary, as we will see, there the transfer operators can be understood as projectors onto the space of so--called physical states, which can be characterized as being in the kernel of (Hamiltonian) constraints. This illustrates the general principle that a path integral with local symmetries acts a projector on a constraint subspace \cite{hartle}. On the other hand, one can construct a projector onto the constraint subspace as a path integral. See  \cite{karim}, in which this is performed for $3D$ gravity, which can be formulated as an $SU(2)-BF$--theory.

The transfer operator $\hat T$ is defined on a Hilbert space $\cal H$, so that the partition function can be written as
\ba\label{t1}
Z&=& \text{tr}_{\cal H}\, \hat T^N  \q .
\ea
Here we assume for simplicity\footnote{See \cite{tent,hoehn} for  a possibility to introduce a slicing of triangulations into hypersurfaces, and a transfer operator based on so--called tent moves.} that we have a hyper--cubical lattice, with one lattice direction designated as time direction, in which we have periodic boundary conditions. The trace $\text{tr}_{\cal H}$ is the trace in a Hilbert space which we will describe in the following. The Hilbert space is then associated to the spatial lattice and can be built as a direct product of Hilbert spaces associated to the spatial edges $e_s$, so that ${\cal H}={\bigotimes}_{e_s} {\cal H}_{e_s}$. As the (configuration) variables associated to the edges are group elements, a Hilbert space associated to an edge will be ${\cal H}_e=\Cl[G]$, i.e. the space of functions on the group, equipped with an inner product (all functions have finite norm as we consider finite groups), which we will give below.

We will work in the coordinate (that is holonomy or connection) representation, and choose as basis states eigenstates $|g\rangle$. For any unitary matrix representation $g\mapsto \rho_{ab}(g)$ of the group $G$ these are common eigenstates of the so called edge holonomy operators $\hat{\rho}(g_e)_{ab}$
\ba\label{t2}
\hat{\rho}(g_e)_{ab}\,|g\rangle&=&\rho(g_e)_{ab} |g\rangle  \q .
\ea
These basis states are orthonormal
\ba\label{t3}
\langle g' \,|\, g\rangle =\prod_{e_s} \delta_G(g'_{e_s}, g_{e_s})
\ea
where $\delta_G$ is the delta function on the group such that $\frac{1}{|G|}\sum_g \delta_{G}(g,g')=1$.  We have also the completeness relation
\ba\label{t4}
\text{Id}_{\cal H} &=& \frac{1}{|G|^{\sharp e_s}} \sum_{g_{e_s}} \,|g\rangle\langle g|   \q .
\ea

We use this resolution of identity $N$ times in (\ref{t1}) to rewrite the partition function as
\ba\label{t5}
Z&=& \text{tr}_{\cal H}\, \hat T^N  \;=\;  \frac{1}{|G|^{\sharp e_s}} \sum_{g_{e_s,n}}  \prod_{n=0}^N \langle g_{n+1}|\,\hat T\,|g_n\rangle \q
\ea
where we introduced $n=0,\ldots, N$ to label the `constant time hypersurfaces' in the lattice. On the other hand we will assume that our partition function is of the form
\ba\label{t6}
Z&=&\frac{1}{|G|^{\sharp e}} \sum_{g_e} \,\prod_f w_f(h_f) \;=\; \frac{1}{|G|^{\sharp e}}  \prod_{n=0}^N \prod_{(f_{s})_{n+1}} w_{f_s}^{\frac{1}{2}}(h_{f_s}) \,\prod_{(f_{t})_n} w_{f_t}(h_{f_t})  \,\prod_{(f_{s})_{n}} w_{f_s}^{\frac{1}{2}}(h_{f_s})   \q .
\ea
Here we introduced weight functions $w_f$ of the holonomies around plaquettes $h_f$, which in the form on the right hand side can also chosen to differ for spatial plaquettes $f_s$ and timelike plaquettes $f_t$. (This is necessary if one wants to obtain the Hamiltonian in the limit of small lattice constant in timelike direction.) Since we are dealing with a gauge theory the weights $w_f$ are class functions, i.e. invariant under conjugation, furthermore we will assume that $w_f(h)=w_f(h^{-1})$, i.e. the weights are independent of the orientation of the face. A weight function can therefore be expanded into characters, which are linear combinations of  matrix elements of representations (in fact characters are just the trace). Hence the weight functions will be quantized as edge holonomy operators.

On the right hand side of (\ref{t6}) we have just split the product into factors labelled by the time parameter $n$. Here we assume that the plaquettes $(f_t)_n$ are the ones between time $n$ and $n+1$.  Comparing the two forms of the partition functions (\ref{t5}) and (\ref{t6}) we can conclude that
\ba\label{t7}
\langle g_{n+1}|\,\hat T\,|g_n\rangle &=&  \,\prod_{f_{s}} w_{f_s}^{\frac{1}{2}}(h_{(f_s)_{n+1}}) \, \, \frac{1}{|G|^{\sharp e_t}}\sum_{g_{e_t}}\prod_{(f_{t})} w_{f_t}(h_{(f_t)_{n}})  \,\,\prod_{f_{s}} w_{f_s}^{\frac{1}{2}}(h_{(f_s)_n})  \q .
\ea

The two outer factors can be easily quantized as multiplication operators, so that we can write
\ba\label{t8}
\hat T &=&  \hat W \hat K \hat W
\ea
with
\ba\label{t9}
\hat W\;=\; \prod_{f_{s}} \hat w_{f_s}^{\frac{1}{2}}(h_{f_s})  \q , \q\q
\langle g_{n+1}  |\, \hat K \,| g_n\rangle  \;=\;  \frac{1}{|G|^{\sharp e_t}}\sum_{g_{e_t}}\prod_{(f_{t})} w_{f_t}(h_{(f_t)_{n}})  \q .
\ea

\begin{figure}[hbt!]
\begin{center}
    \psfrag{t1}{$g_{(e_s)_n}$}
    \psfrag{t2}{$g_{(e_t)_n}$}
    \psfrag{t3}{$g_{(e_s)_{n+1}}$}
    \psfrag{t4}{$g_{(e'_t)_n}$}
    \includegraphics[scale=0.55]{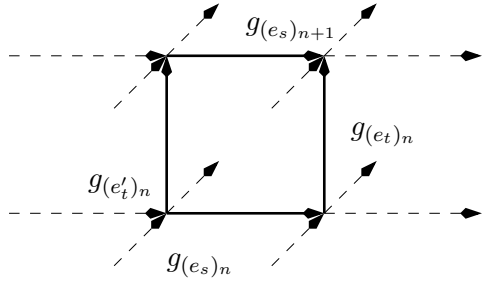}
    \caption{A time like plaquette in a cubical lattice. The group elements at the time like edges can be understood to act as gauge transformations on the vertices either at time steps $n$ or $n+1$. Integration over the group elements assoicated to the time like edges enforces therefore (via group averaging) a projector onto the gauge invariant Hilbert space.}\label{plaques}
\end{center}
\end{figure}

To tackle the operator $\hat K$, note that the group elements associated to the timelike edges appearing in the plaquette weights
\ba\label{t10}
w_{f_t}\big(\, g_{(e_s)_n} g_{(e_t)_n} g^{-1}_{(e_s)_{n+1}} g^{-1}_{(e'_t)_n} \,\big)
\ea
can be understood as acting as gauge transformations associated to the vertices of the spatial lattice, see figure \ref{plaques}.

That is we define operators $\hat \Gamma(\gamma),\, \gamma\in  G^{\sharp v_s} $ by
\ba\label{t11}
\hat \Gamma(\gamma)\,|g\rangle &=&  |\gamma^{-1}\rhd  g\rangle
\ea
where $(\gamma\rhd g)_e = \gamma_{s(e)} g_e \gamma^{-1}_{t(e)}$ and $s(e),t(e)$ are the source and target vertex of the edge $e$, respectively. The operators $\hat \Gamma$ generate gauge transformations as
\ba\label{t12}
\langle g| \hat \Gamma(\gamma)|\psi\rangle \;=\; \langle \gamma \rhd g |\psi \rangle \;=\; \psi( \gamma \rhd g) \q .
\ea

Now we can see the plaquette weight in (\ref{t10}) as either a wave function at time $n$ or a wave function at time $n+1$ on which the group elements associated to the timelike edges act as a gauge transformation. The sum in (\ref{t9}) over the group elements $g_{e_t}$ then induces an averaging over all gauge transformations, i.e. a projection onto the space of gauge invariant states. Hence we can write
\ba\label{t13}
\hat K = \hat P_G \hat K_{0}  = \hat K_{0} \hat P_G
\ea
where
\ba\label{t14}
\hat P_G \;=\; \frac{1}{|G|^{\sharp v_s} }\sum_\gamma \hat \Gamma(\gamma) \q , \q\q\langle g_{n+1}|\,\hat K_0\, |g_n\rangle \;=\;\prod_{f_t} w_{f_t}(g_{(e_s)_n} g^{-1}_{(e_s)_{n+1}})  \q .
\ea
The operator $\hat P_G$ is a projector, that is $\hat P^2_G=\hat P_G$. One can easily show that $\hat \Gamma(\gamma)\hat P_G=\hat P_G \hat \Gamma(\gamma)=\hat P_G$ for any $\gamma \in G^{\sharp v_s} $, hence it projects onto the space of gauge invariant functions.  As the operator $\hat W$ is made up from gauge invariant plaquette couplings it commutes with $\hat P_G$, so that we can write
\ba\label{t15}
\hat T= \hat P_G \hat W \hat K_0 \hat W \hat P_G  \q .
\ea

Here we see an example for a general mechanism, namely that the transfer operator for a partition function with local gauge symmetries acts as a projector onto the space of gauge invariant states. We can characterize such gauge invariant states as being annihilated by constraint operators. In the case of the usual lattice gauge symmetries these constraints are the Gau\ss~constraints, which we will discuss later--on.

To discuss the action of $K_0$ we  introduce first the spin network basis, in which $\hat K_0$ will be diagonal.

\subsection{Spin network basis}

So far we encountered the holonomy operators, i.e. the multplication operators $\hat\rho(g_e)_{ab}$ in the configuration representation. The conjugated operators are the electric fields or fluxes, which act as (matrix) multiplication operators in the spin network basis. This is a convenient basis to discuss the remaining part $K_0$ of the transfer operator.

To obtain the spin network basis \cite{rovellibook,thomasbook} we just need to use that every function on the group can be expanded as  a sum over the matrix elements $\rho(\cdot)_{ab}$ of all irreducible unitary representations $\rho$. For Abelian groups all irreducible representations are one--dimensional, hence $a,b=0$ and we obtain the discrete Fourier transform (\ref{za3}). In the general case of non-Abelian groups we define spin net work states $|\rho,a,b\rangle$ by
\ba\label{tb1}
\prod_e \sqrt{\dim \rho_e}\, \rho_e(g_e)_{a_e b_e} &=& \langle g | \rho,a,b\rangle  \q .
\ea
\noindent which are orthonormal, i.e. $\langle\rho,a,b|\rho',a',b'\rangle=\delta_{\rho\rho'}\delta_{aa'}\delta_{bb'}$

In the spin network basis we can easily express the left $L_e(h)$ and right translation operators $R_e(h)$. These are the conjugated operators to the holonomy operators \ref{t2}. To simplify notation we will just consider the Hilbert space and states associated to one edge and omit the edge subindex. We define
\ba\label{tb2}
\hat L(h)\,|g\rangle\;=\; |h^{-1} g\rangle  \q , \q\q  \hat R(h)\,|g\rangle\;=\; | g h\rangle \q .
\ea
In the spin network basis
\ba\label{tb3}
\langle g|\hat L(h) |\rho,a,b\rangle\;&=&\; \langle hg| \rho,a,b\rangle \;=\; \sqrt{\dim \rho}\;\rho(hg)_{ab}\\[5pt]\nonumber
\;&=&\;\sqrt{\dim \rho}\;\rho(h)_{ac}\,\rho(g)_{cb}\;=\;\rho(h)_{ac} \langle g|\rho,c,b\rangle \q .
\ea
where we sum over repeated indices. That is
\ba\label{tb4}
\hat L(h) |\rho,a,b\rangle\;=\; \rho(h)_{ac}|\rho,c,b\rangle \q, \q\q  \hat R|\rho,a,b\rangle\;=\; |\rho,a,c\rangle \, \rho(h^{-1})_{cb} \q .
\ea

The remaining factor $K_0$ of the transfer operator factorizes over the edges and it is straightforward to check that its acts on one edge as
\ba\label{tb5}
\hat K_0 \;=\; \frac{1}{|G|}\sum_h w_{f_t}(h) \,\hat L(h) \;=\;  \frac{1}{|G|}\sum_h w_{f_t}(h) \,\hat R(h) \q .
\ea
(Here one has to use that $w_{f_t}$ is a class function and invariant under inversion of the argument.) On a spin network state we obtain
\ba\label{tb6}
\hat K_0\,|\rho,a,b\rangle &=& \frac{1}{|G|} \sum_h w_{f_t}\, \rho(h)_{ac}|\rho,c,b\rangle \nn\\
&=&\frac{1}{|G|} \sum_h  \sum_{\rho'} c_{\rho'} \chi_{\rho'}(h) \rho(h)_{ac} \,|\rho,c,b \rangle \nn\\
&=& \sum_\rho  \frac{1}{\dim\rho} c_\rho \,|\rho,a,b\rangle
\ea
where in the second line we used that a class function can be expanded into characters $\chi_\rho$ and in the third the orthogonality relation
\ba\label{tb7}
\frac{1}{|G|} \sum_h \chi_{\rho'}(h) \rho(h)_{ac} &=& \frac{1}{\dim\rho} \,\delta_{\rho,\rho'} \,\delta_{ac}   \q .
\ea
The expansion coefficients $c_\rho$ are given by
\ba
c_\rho = \frac{1}{|G|} \sum_h w_{f_t}(h) \bar \chi_\rho(h)
\ea

That is $K_0$ acts diagonal in the spin network basis and as the eigenvalues only depend on the representation labels $\rho$ it commutes with left and right translations.  For Yang--Mills theory (with Lie groups) in the limit of continuous time one obtains for $K_0$ the Laplacian on the group  \cite{kogut_susskind, smit}.


\subsection{Gauge invariant spin nets}\label{gi}

Given a spin network function $\psi$, which can be regarded as function $\psi(h_e)$ of holonomies associated to edges,  where $h\in G^{\sharp e}$ is a connection, and a gauge transformation $\gamma\in G^{\sharp v}$ an assignment of group elements to vertices of the network, then the gauge transformed spin network function is given by
\begin{eqnarray}
\hat\Gamma(\gamma)\psi(h_e)\;:=\;\psi(\gamma_{s(e)}h_e \gamma_{t(e)}^{-1})
\end{eqnarray}

\noindent where $s(e)$ and $t(e)$ are the source and the target vertices of the edge $e$. A gauge transformation can therefore be expressed as a linear operator in terms of $\hat L$ and $\hat R$, as shown in the last section. Decomposing the spin network function into the basis $|\rho_e,a_e,b_e\rangle$, i.e.
\begin{eqnarray}
\psi(h_e)\;=\,\sum_{\rho_e,a_e,b_e}\tilde\psi_{\rho_e,a_e,b_e}|\rho_e,a_e,b_e\rangle
\end{eqnarray}

\noindent one can readily see that the condition for a function $\psi$ to be invariant under all gauge transformations $\alpha_g$ translates into a condition for the coefficients $\tilde\psi_{\rho_e,a_e,b_e}$, namely
\begin{eqnarray}
\tilde\psi_{\rho_e,a_e,b_e}\;=\;\prod_v\iota^{(v)}_{a_{e_1}\ldots,b_{e_n}}
\end{eqnarray}

\noindent where the product ranges over vertices $v$, and each tensor $\iota^{(v)}$, which has the indices $a_e$ for each edge $e$ outgoing and $b(e)$ for each edge $e$ incoming $v$, is an invariant element of the tensor representation space
\begin{eqnarray}
\iota^{(v)}\;\in\;\text{Inv}\left(\bigotimes_{v=s(e)}V_{\rho_e}\otimes\bigotimes_{v=t(e)}V_{\rho_e}^*\right)
\end{eqnarray}

\noindent i.e. an intertwiner between the tensor product of incoming and outgoing representations for each vertex. An orthonormal basis on the space of gauge-invariant spin network functions therefore corresponds to a choice of orthonormal intertwiner in each tensor product representation space for each vertex, where the inner product is the tensor product of the inner products on each $V_\rho$, $V_\rho^*$

Note that neither the holonomies $\hat\rho(h_e)_{ab}$ nor left- and right translations are gauge-invariant, therefore they map gauge-invariant functions to non-gauge-invariant ones. However, they can be combined into gauge-invariant combinations, such as the holonomy of a Wilson loop within a net.

\subsection{Local constraint operators and physical inner product}

We have seen that for a general gauge partition function of the form (\ref{t6}) the transfer operator (\ref{t15})
\ba\label{tc1}
\hat T= \hat P_G \hat W \hat K_0 \hat W \hat P_G
\ea
incorporates a projection $\hat P_G$ onto the space of gauge invariant states. This is a general feature of partition functions with gauge symmetries \cite{hartle}. Indeed in quantum gravity one attempts to construct a projector onto gauge invariant states by constructing an appropriate partition function.

As has been discussed in section \ref{symm} the $BF$--models enjoy a further gauge symmetry, the translation symmetries (\ref{z7}). (A generalization of these symmetries holds also for non-Abelian groups.) Hence we can expect that another projector will appear in the transfer operator. Indeed it will turn out that the transfer operator for the $BF$--models is just a combination of projectors.

This is straightforward to see as for the $BF$--models the plaquette weights are given by $w_f(h)=\delta_G(h)$, so that
\ba\label{tc2}
\hat W &=& \prod_{f_s} \big( \hat \delta_G(h_{f_s})\big)^\frac{1}{2}   \q .
\ea
Here one might be worried by taking the square roots of the delta functions, however we are on a finite group. The operator $\hat W^2$ is diagonal in the basis $|g\rangle$ and has only two eigenvalues: $q^{\sharp f_s}$ on states where all plaquette holonomies vanish and zero otherwise. The square root of $\hat W^2$ is defned by taking the square root of these eigenvalues.

Hence $\hat W$ is proportional to another projector $\hat P_F$, which projects onto the states for which all the plaquette holonomies are trivial, that is states with zero (local) curvature.

The final factor in the transfer operator $\hat T$ is $\hat K_0$ which according to the matrix element of $\hat K_0$ in (\ref{t14}) (putting $w_f(h)=\delta_G(h)$) is proportional to the identity
\ba\label{tc3}
\hat K_0&=& |G|^{\sharp e_s}\,\, \text{Id}_{\cal H}
\ea
where the numerical factor arises due to our normalization convention for the group delta functions, which amounts to $\delta_G(\text{id})=|G|$. Hence the transfer operator for $BF$--theory is given by
\ba\label{tc4}
\hat T  \;=\; |G|^{\sharp e_s+ \sharp f_s} \,\,\hat P_G  \hat P_F \;= \;|G|^{\sharp e_s+ \sharp f_s}\,\, \hat P_F  \hat P_G  \q
 \ea
and since for the projectors we have $\hat P_G^N=\hat P_G,\hat P_F^N=\hat P_F$ the partition function simplifies to
\ba\label{tc4a}
Z&=&\text{tr}_{\cal H} \, \hat T^N \;=\; |G|^{N(\sharp e_s+ \sharp f_s)} \,\text{tr}_{\cal H}  \hat P_G \hat P_F\;=:\, |G|^{N(\sharp e_s+ \sharp f_s)} \,\text{dim} (\,{\cal H}_{phys})  \q .
\ea
The projection operators in the transfer operator ensure that only the so called physical states $\psi\in {\cal H}_{phys}$ are contributing to the partition function $Z$. These are states in the image of the projectors (we will call the corresponding subspace  ${\cal H}_{phys}$) and can be equivalently described to be annihilated by constraint operators, which we will discuss below.

The objective in canonical approaches to quantum gravity \cite{rovellibook, thomasbook} is to characterize the space of physical states according to the constraints given by general relativity, which follow from the local symmetries of the theory. (The quantization of these constraint in a consistent way is highly complicated \cite{thiemannH}.) Indeed also in general relativity, as well as in any theory which is invariant under reparametrizations of the time parameter, one would expect that the transfer operator (or the path integral) is given by a projector, so that the property $\hat T^2 \sim \hat T$ would hold. \footnote{In the limit of taking the discretization in time direction to the continuum we would obtain the same projector as for finite time discretization. Indeed the projectors already encode for finite time steps the (Hamiltonian) constraints, which are the generators of time translations (time reparametrization symmetry).}This would also imply a certain notion of discretization independence, namely, the independence of transition amplitudes on the number of time steps used in the discretization, see also the discussion in \cite{seb} on the relation between reparametrization symmetry in the time parameter and discretization independence. For a lattice based on triangulations one can introduce a local notion of evolution, the so called tent moves \cite{tent, hoehn}. These moves evolve just one vertex and the adjacent cells. Local transfer operators can be associated to these tent moves. These would evolve the spatial hypersurface not globally, but locally, and would allow to choose some arbitrary order of the vertices to evolve. In this way one can generate different slicings of a triangulation as well as different triangulations. The condition that the transfer operator associated to a tent move is given by a projector would then implement an even stronger notion of triangulation independence.

However reparametrization invariance, which would lead to such transfer operators, is usually broken by  discretizations already for one--dimensional toy models. For (coarse graining and renormalization) methods to regain this symmetry, see \cite{seb,coarse1,perfect}. In the case of gravity, or more generally non--topological field theories, these methods will however lead to non--local couplings for the partition functions \cite{coarse1}, for which one would have to modify the transfer operator formalism.

Furthermore one has to construct a physical inner product on the space of the physical sates. This process is quite trivial in the case considered here, as we are considering finite groups and all the projectors are proper projectors. Hence the physical states are  proper states in the (so--called) kinematical Hilbert space ${\cal H}$ and the inner product of this space can be taken over for the physical states. In contrast, already for $BF$--theory with (compact) Lie--groups the translation symmetries lead to non--compact orbits. This leads to physical states which are not normalizable with respect to the inner product of the kinematical Hilbert space. For the intricacies of this case (with $SU(2)$) see for instance \cite{karim}. A further complication for gravity is the very complicated non--commutative structure of the constraints \cite{master}.

\begin{figure}[hbt]
\begin{center}
    \includegraphics[scale=0.3]{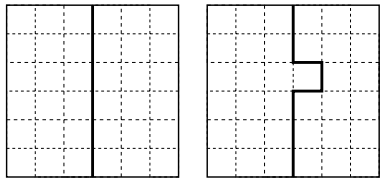}
    \caption{Two states in the spin network basis for $\Zl_2$ which are equivalent under the physical inner product. The thick edges carry non--trivial representations. The right state can be obtained from the left state by applying a plaquette stabilizer. }\label{spinnetd}
\end{center}
\end{figure}

Let us summarize the projectors into $\hat P=\hat P_G\, \hat P_F$. Also in the case of generalized projectors a physical inner product  can be defined \cite{RAQ,thomasbook} between equivalence classes of kinematical states labelled by representatives $\psi_1,\psi_2$ through
\ba\label{tc5}
\langle \psi_1| \psi_2 \rangle_{phys} &:=& \langle \psi_1|\hat P | \psi_2 \rangle  \q .
\ea
Kinematical states which project to the same (physical) state define the same equivalence class. This leads for instance to an identification of the two states (for the group $\Zl_2$) in figure \ref{spinnetd}. This is analogous to the action of the spatial diffeomorphism group in ($3D$ and $4D$)  loop quantum gravity, see the discussion in \cite{karim} and reflects the concept of abstract spin foams, whose amplitude does not depend on the embedding into the lattice, in section \ref{ising}. Indeed any two states which can be deformed into each other by applying the so--called stabilizer operators introduced below in (\ref{tc6},\ref{tc12}) are equivalent to each other. The reason is that the physical Hilbert space is the common eigenspace to the eigenvalue $1$ with respect to all the stabilizer operators. Hence any part of a state that undergoes a non--trivial change if a stabilizer is applied, will be projected out in the physical inner product.

Another problem in quantum gravity is then to find (quantum) observables \cite{observ} that are well defined on the physical Hilbert space. In the case of gravity these Dirac observables are very hard to obtain explicitly, even classically there are almost none known . In particular, such Dirac observables have to be non--local \cite{torre}. If one interprets a quantum gravity model as a statistical system a related task is to find a well--defined order parameter characterizing the phases. Expectation values of gauge variant observables may be projected to zero under the physical inner product \cite{elitzur}: As in our case we work with finite dimensional Hilbert spaces and the physical Hilbert space is just a subspace of the kinemtaical one, there is a construction principle for physical observables. For any operator $\hat O$ on the kinematical Hilbert space one can consider $\hat P \hat O \hat P$, which preserves the physical Hilbert space and corresponds to a 'gauge averaging' of $\hat O$. However many observables constructed in this way will turn out to be constants. For $BF$--theory  observables can be constructed by considering a product of the stabilizer operators discussed below along non--contractable loops \cite{kitaev, louapre}. The resulting observables are non--local, and the cardinality of an set of independent operators (equivalently the dimension of the physical Hilbert space) just depends on the topology of the lattice and not on its size.

We will now discuss the stabilizer operators which characterize the physical states. In topological quantum computing these are known as stabilizer conditions \cite{kitaev}.

We have considered the operators $\hat \Gamma(\gamma)$ in (\ref{t11}) that implement a gauge transformation according to the assingements $(\gamma)_{v_s}$ of group elements to the vertices of the (spatial sub--) lattice. These can be easily localized to the vertices $v_s$ of the spatial lattice by choosing the gauge parameters $\gamma$  (an assignement  of one group element to every vertex in the spatial lattice) such, that all group elements are trivial except for one vertex $v_s$. We will call the corresponding operators $\Gamma_{v_s}(\gamma)$ where now $\gamma$ denotes an element in the gauge group $G$ (and not in the direct product $G^{\sharp v_s}$). These can be expressed by the left and right translation operators (\ref{tb2})
\ba\label{tc6}
\hat \Gamma_{v_s}(\gamma) \,|g\rangle &=& \prod_{e_s:\,s(e_s)=v_s} \hat L_{e_s}(\gamma) \prod_{e'_s:\,t(e'_s)=v_s} \hat R_{e'_s}(\gamma) \,|g\rangle  \q .
\ea
Physical states  $|\psi\rangle$ have to satisfy\footnote{Here $\hat \Gamma_{v_s}$ is a unitary operator, so that the condition on the physical states is in the form of a so--called stabilizer. Alternatively the condition can be expressed by self--adjoint constraints $\hat C_{v_s}(\gamma)=i(\Gamma_{v_s}(\gamma)-\Gamma_{v_s}(\gamma^{-1}))$ for group elements $\gamma$ with $\gamma \neq \gamma^{-1}$. Otherwise define $C_{v_s}(\gamma)=\Gamma_{v_s}(\gamma)-\text{Id}_{\cal H}$. Physical states are then annihilated by the constraints.}
\ba\label{tc7}
\hat \Gamma_{v_s}(\gamma) |\psi\rangle &=& |\psi\rangle
\ea
for all vertices $v_s$ and all group elements $\gamma$. As the operators $\hat \Gamma$ define a representation of the group we can restrict to the elements of a generating set of the group.

This suggest to consider the action of these (star) operators in the spin network basis:
\ba\label{tc8}
\hat \Gamma_{v_s}(\gamma) \,|\rho,a,b\rangle &=&  \prod_{e:\,s(e)=v_s} \rho_e(\gamma)_{a_e c_e} \prod_{e':\,t(e')=v_s} \rho_{e'}(\gamma^{-1})_{d_{e'} b_{e'}} \,\,| \rho, (c_e,a_{e'},a_{e''}), (b_e,d_{e'},b_{e''})\rangle \q\q\q\q
\ea
where on the right hand side edges $e$ are the ones starting at $v_s$, $e'$ those ending at $v_s$ and $e''$ are all the remaining edges in the spatial lattice. From this expression one can conclude that the spin network states transform at $v_s$ in a tensor product of representations
\ba\label{tc9}
\rho_{v_s} &=& \bigotimes_{e:\,s(e)=v_s} \rho_{e}  \,\,\otimes \,\,  \bigotimes_{e':\,t(e')=v_s} \rho^*_{e'}
\ea
\noindent
where $\rho^*$ is the dual representation to $\rho$. Gauge invariant states can be constructed by considering the projections of these representations to the trivial one, or equivalently by contracting the matrix indices of the states  with the intertwiners between the tensor product of `incoming' representation and the tensor product of 'outcoming' representations at the vertices, see section \ref{gi}.

For the Abelian groups $\Zl_q$ the irreducible representations are one--dimenionsional, hence we can omit the indices $a,b$ in the spin network basis. The tensor product of representations (\ref{tc9}) is also one dimensional and equal to the trivial representation, provided
\ba\label{tc10}
\sum_{e:\,s(e)=v_s} k_e &=& \sum_{e':\,t(e')=v_s}  k_{e'}
\ea
for the representation labels $k_e,k_{e'}$ of the outgoing and incoming edges at $v_s$ respectively. Hence we recover the Gau\ss~constraints from the spin foam representation (\ref{y4}). Here these Gau\ss~constraints appear as based on vertices as opposed to edges as in (\ref{y4}). But the spatial vertices $v_s$ represent the time like edges and the Gau\ss~constraints in the canonical formalism are just the Gau\ss~constraints associated to the time like edges in the spin foam representation (\ref{y4}).

Let us turn to the other projector $\hat P_F$. It maps to states for which all the spatial plaquettes $f_s$ have trivial holonomies. Hence the conditions on physical states can be written as
\ba\label{tc11}
\hat F_{f_s,a,b}^\rho |\psi\rangle&=&\delta_{ab}|\psi\rangle
\ea
where we have introduced the plaquette operators (corresponding to the flatness constraints)
\ba\label{tc12}
\hat F_{f_s,a,b}^\rho = \left(\vec\prod_{e\subset f_s} \hat \rho(g_{e}^{o(f_s,e)}) \right)_{ab}  \q .
\ea
Here $\rho$ should be a faithful representation, otherwise one might find, that also states with local non--trivial holonomies satisfy the conditions (\ref{tc11}), see for instance the discussion in \cite{perez}. On the other hand this can be taken as one possible generalization of the model. For non--Abelian groups the plaquette operators additionally depend on the choice of a vertex adjacent to the face, at which the holonomy around the face starts and ends. However the conditions (\ref{tc11}) do not depend on this choice of vertex: if a holonomy around a face is trivial for one choice of vertex it will be trivial for all other vertices in this face, as these holonomies just differ by a conjugation.

$BF$--theory in any dimension is a topological field theory, that is there are only finitely many physical degrees of freedom depending on the topology of space. Consequently the number of physical states, or equivalently of equivalence classes of kinematical states in the sense (\ref{tc5}) is finite and does not scale with the lattice volume.

There are a number of generalizations one could consider. First one can couple matter in the form of defects to the models. In $3D$ $SU(2)$--$BF$ theory corresponds to a first order formulation of $3D$ gravity.  
 Point-like particles can be coupled to the model, see for instance \cite{louapre} and lead to the violation of the flatness constrainst (\ref{tc12}) (coupling to the mass of the particles) and to the violation of the Gau\ss~constraints (\ref{tc7}) (coupling to the spin of the particles) at the position of the particle. The defects can be understood as changing the topology of the spatial lattice, i.e. to change its first fundamental group. To have the same effect in, say four dimensions, one needs to couple strings, see \cite{baezS}.

In the context of quantum computing \cite{kitaev} one does not necessarily impose the conditions (\ref{tc7}) and (\ref{tc12}) as constraints but as characterizations for the ground states of the system. Elementary excitations or quasi particles are states in which either the flatness or the Gau\ss~constraints are violated. These excitations appear in pairs (at least for Abelian models \cite{bombin}) and can be created by so--called ribbon operators \cite{kitaev, bombin, kadar}, which in the context of the proper particle models in $3D$ gravity are related to gauge invariant (Dirac) observables \cite{freidel_u}. For further generalizations based on symmetry breaking from $G$ to a subgroup of $G$, see for instance \cite{bombin}.

In $4D$ gravity is not topological, rather there are propagating degrees of freedom.  Similar to the discussion for the partition functions in \ref{symm} one can try to construct new models which are nearer to gravity by breaking down the symmetries of $BF$--theory. In $3D$ the flatness constraints are defined on the plaquettes of the lattice. Constraints act also as generators of gauge transformations (indeed the conditions (\ref{tc7},\ref{tc12})  impose that physical states have to be gauge invariant) and the flatness constraints can be interpreted as translating the vertices of the dual lattice in space time \cite{zapata,dittrichr,bahrdittrich1}, which can be seen as an action of a diffeomorphism. In $4D$ the same interpretation holds only for some exceptionally cases corresponding to special lattices that do not lead to space time curvature, see the discussion in \cite{dittrichryan}.  In general the flatness constraints rather generate translations of the edges of the dual lattice \cite{bonzom4d}.

A possible generalization leading to gravity--like models is therefore to replace the flatness conditions (\ref{tc11}) based on plaquettes with some contractions of these conditions, such that these new conditions are based on the $3$--cells, i.e. cubes in a hyper--cubical lattice. This would correspond to the contractions of the form $FEE$ (Hamiltonian constraints) and $FE$ (diffeomorphism constraints) in the (complex) Ashtekar variables of the curvature $F$ with the electric flux variables $E$ \cite{ashtekar, rovellibook, thomasbook}. The difficulty in constructing such models is consistency of the constraint algebra, i.e. to find constraints that form a closed algebra. However to consider such models for finite groups should be much simpler than in the case of full gravity. Even if such consistent constraint algebras cannot be found, physical Hilbert spaces could be constructed, for instance with the techniques in \cite{master, uniform}.

Furthermore it will be illuminating to derive the transfer operators for the constrained models introduced in section \ref{constra}, as in the full theory the relation between the covariant models and Hamiltonians in the canonical formalism is still open \cite{relation}. To this end a connection representation \cite{ oecklpf,pfeifferc} can be employed.



\section{Group field theory (GFT)}
\label{gft}

We have seen in earlier sections that the spin foam formalism is the natural description of the high temperature expansion for random lattices. This is by no means the only instance that random lattices have been used to analyze the properties of dynamical systems.  As we shall see, group field theories \cite{deppet} describe weighted sums over $n$-dimensional random lattices and have enough freedom to capture the dynamics of many dynamical systems. They attempt to capture the degrees of freedom of local spin foam models stemming from a sum over cellular complexes. 

In the 2-dimensional case, they are known by the more familiar moniker of (random) matrix models \cite{difran, davidrev}.  These are well-studied in a variety of fields and can describe a multitude of different scenarios including, but not limited to, 2d gravity with cosmological constant \cite{grav}, 2d Ising/Potts models \cite{ising}, 2d Yang-Mills theories \cite{yang}, certain string theories \cite{string}, the enumeration of virtual knots and links \cite{tangle}, the list goes on.  Moreover, they have inspired the development of a plethora of useful techniques to solve and analyse statistical ensembles of random matrices \cite{mehta}  such as the topological expansion \cite{thooft, brezin}, the eigenvalue method \cite{eigen}, the double-scaling limit, the method of orthogonal polynomials \cite{poly} and the character expansion \cite{char}, to name just a few.  The more recent label of GFT is to higher-dimensional random lattice theories as matrix model is to 2d random lattice theories and they share the same motivation.  The ambition is to develop a similar technology for GFTs and the development of an apparatus to count the degree of divergence \cite{divcount} and later a topological expansion has had its first success \cite{gurautop}, but tools to extensively solve these models are still missing.

To say it in yet another way, GFTs are a natural tensor generalization of matrix models. They generate a weighted sum over n-dimensional cell complexes in the perturbative expansion of the free energy.  They have garnered increasing attention from the quantum gravity community, with the advent of the spin foam formalism.  Not only do spin foam diagrams naturally arise as the Feynman graphs of GFTs \cite{reisen} but inherent in all current spin foam models  for 4d quantum gravity is  a truncation of the degrees of freedom and a manifest loss of diffeomorphism symmetry.   It is with the hope of re-obtaining these lost properties (e.g. perhaps in a the continuum limit regime) that one analyses these higher-dimensional theories. Tensor models first appeared in the quantum gravity literature with the work of Ambjorn et al. \cite{ambj}.  Later they were generalized to genuine multi-variable field theories over compact Lie groups, such as SU$(2)$ (or more generally the internal local symmetry group of the $n$-dimensional gravity theory in question). These are known as Boulatov-Ooguri field theories \cite{bou, oog, depfre, fra}.     With this further generalization, one does not think of these models as generating discrete cell complexes with unit $n$-volume cells.  Rather the field arguments carry the local dynamical geometrical information for the cell complex and thus each Feynman graph already contains a restricted sum over geometries, at least in theory. The sum over graphs completes the sum over geometries (and topologies).  Moreover, higher-dimensional finite group models deserve more study since we shall see that they are instrumental in describing certain matter degrees of freedom.

To commence, let us detail some of the general set-up.

\subsection{General set-up}

While one might hope to proceed at the outset with the most generic definition possible, this is probably not the wisest course of action.  On the other hand, we shall not be too specific either, but try only to put forth just a flavour of the theory at this stage.  With this in mind, let us define a somewhat, but not totally, general GFT \cite{petro}.  The fundamental dynamical object is the field $\phi: G^{\times n}\rightarrow \Rl$; a field on $n$ copies of a group $G$.  Schematically, the action takes the form:
\be\label{gft01}
\begin{array}{rcl}
\cs[\phi] &=&\dsty \int_{G^{\times 2n}}\left[ \prod_{a=1}^n dg_a\; d\hat g_a\right] \phi(g_1, \dots, g_n) \;\ck(\{g_a \hat g_a^{-1}\})\; \phi(\hat g_1,\dots,\hat g_n)\\[0.5cm]
 &&\dsty- \frac{\lambda}{n!} \int_{G^{\times n(n+1)}} \left[\prod_{a,b = 1:b\neq a}^{n+1} dg_{a,b}\right] \left[\prod_{a = 1}^{n+1} \phi(g_{a,a-1},\dots,g_{a,a-n})\right]\;\cv(\{g_{a,b}^{\phantom{-1}}\,g_{b,a}^{-1}\}).\\[0.5cm]
 \end{array}
\ee
We use a normalized measure on $G$, while $\ck$ and $\cv$ are the kinetic and vertex operators.  As usual, they are instrumental in the construction of Feynman amplitudes.  One should pay attention to the peculiar coupling of the arguments in \eqref{gft01}, especially in the vertex operator.  As a field theory on $G^{\times n}$, the vertex interaction is highly non-local. But therein lies the power of the GFT framework. When one calculates the perturbative expansion of the free energy, one can in fact view the Feynman graphs as $n$-dimensional cell complexes $\Gamma$. What is more, for the $\phi^{n+1}$ interaction above, this cell-complex is topologically dual to an $n$-dimensional triangulation $\Delta$.

\begin{figure}[h]
\centering
\includegraphics[width = 8cm]{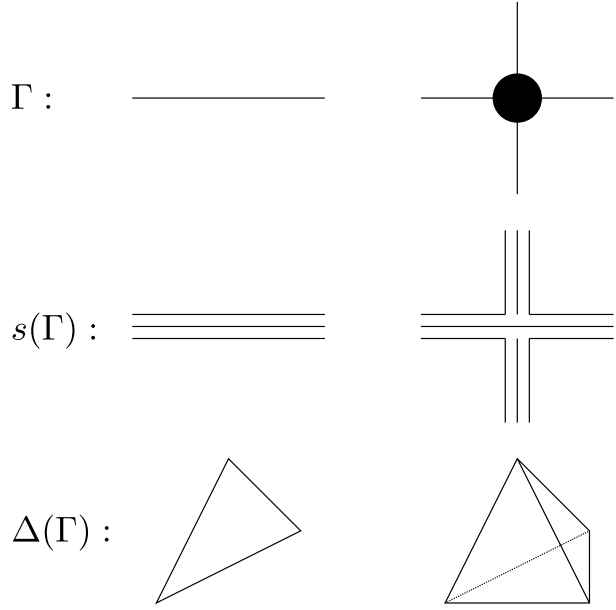}
\caption{\label{fatdraw} For the case $n=3$, we have drawn the fundamental components for the Feynman diagram $\Gamma$, the strand diagram $s(\Gamma)$ and the dual n-simplex $\Delta$.}
\end{figure}

To get an idea of how this might be the case, let us start with a generic Feynman diagram occurring in the expansion of the free energy:
\be
\cf = \log\cz = \int \cd\phi\;  e^{-S[\phi]} = \sum_{\Gamma} \frac{1}{|Aut[\Gamma]|}\cz_{\Gamma} \q\textrm{where}\q \cz_{\Gamma} = \lambda^{\#v} \int \mu_\Gamma(G) \prod_{e\in\Gamma}\cp \prod_{v\in\Gamma} \cv
\ee
and $\mu_\Gamma(G)$ is shorthand for the measure on $\Gamma$, $|Aut[\Gamma]|$ is the order of the discrete automorphism group of the graph and $\cp = \ck^{-1}$ is the propagator.  To explain, let us make use of Fig. \ref{fatdraw}. The diagram $\Gamma$ is simply an $n+1$-valent graph with components given on the first line of the diagram. But due to the non-local nature of the vertex operator, we can define a related {\it strand diagram} $s(\Gamma)$ which captures more accurately the richness of the information encoded in the theory. (Strand diagrams are also known as a fat or ribbon graphs.)  For the strand diagram, we draw the field rather as a collection of $n$ strands.  The coupling described in $\cv(\{g_{a,b}^{\phantom{-1}}\,g_{b,a}^{-1}\})$ is drawn as a re-routing of the strands within the vertex interaction.
The vertex we chose in \eqref{gft01} contains enough information to be interpreted as the dual 2-skeleton to an $n$-simplex.  In this dual picture  the field represents a $(n-1)$-simplex. Its $n$ arguments are the $(n-2)$-simplices on the boundary of the $(n-1)$-simplex.  The vertex re-routing identifies $(n-1)$-simplices along their $(n-2)$-simplices so as to form an n-simplex.  Thus, the interaction may be equivalently seen as generating elementary $n$-simplices.  The kinetic term simply glues these $n$-simplices together by identifying a pair of $(n-1)$-simplices on the respective boundaries.  To sum up, the Feynman diagrams with their amplitudes $\cz_\Gamma$ are rich enough in content to deserve a strand diagram, which in turn may be viewed as the dual to an n-dimensional simplicial complex.  One then has an arena within which to frame many of the interesting questions of lattice theories.

 Obviously, there are a number of ways that one can generalize the formalism above:  
\begin{itemize}
\item It is in fact rather rare that one chooses a real map.  Often, one takes a complex field $\phi: G^{\times n}\rightarrow \Cl$ , but in principle there are very few restrictions.  In fact, it is convenient to view the field as carrying a representation of some other group $H$ (finite or compact).   We shall see that this allows one to incorporate additional degrees of freedom, interpretable as matter. 
\item One could also include more than one field.  This is to a certain extent a sub-case of the previous point but a specific type known as colouring has attracted a large amount of interest \cite{colour}.  It has been shown that in the $n$-dimensional model \eqref{gft01}, if one has $n+1$ fields $\phi_i:G^{\times n}\rightarrow \Cl$, $(i\in \{1,\dots,n+1\})$, and a specific type of interaction Lagrangian:
\be
\cl_{vertex} =  \left[\prod_{a = 1}^{n+1} \phi_a(g_{a,a-1},\dots,g_{a,a-n}) - \prod_{a = 1}^{n+1} \bar\phi_a(g_{a,a-1},\dots,g_{a,a-n})\right] \;\cv(\{g_{a,b}^{\phantom{-1}}\,g_{b,a}^{-1}\}),
\ee
one has much more control over the topology of the discrete structures generated in the perturbative expansion.  For $n=2$, one is guaranteed to generate triangulations corresponding to oriented connected Riemann surfaces (although, in this case colouring is not required as long as the field is complex).  Even with colouring, it is no longer the case in higher-dimensional examples that the triangulation corresponds to a manifold.  In $n = 3$ for example, one gets not just manifolds, but also objects known as pseudo-manifolds. The benefit of having a coloured model is that it removes those pseudo-manifolds from the sum that have been shown to contribute more heavily than manifolds to the free energy \cite{lost}.  

\item Lastly, but certainly not leastly, the presence or absence of symmetries plays a prominent role in GFTs.  Usually, these symmetries come in a rather idiosyncratic fashion.  Some models maintain a symmetry under the action of the permutation group on the arguments of the field.   Positing this symmetry serves to increase the number of topologies arising in the sum and/or alters the relative weight of existing configurations. 

In matrix models, anti-symmetry or hermiticity (of the matrix) is a desirable property from the point of view of solvability via the method of eigenvalues.  There are higher-dimensional versions, which have not been linked as yet to methods of solution.

By far the most important symmetry in $n$-dimensional GFT is the so-called diagonal symmetry under the action of the group.   It is these symmetry generators that often truly parameterize the dynamical degrees of freedom of the model at the Feynman diagram level.

\end{itemize}

\subsection{Matrix models}

\subsubsection{2d gravity with cosmological constant}

Consider an Hermitian $N\times N$ matrix: $\phi_{g_ag_b}$ such that $(\phi^\dagger)_{g_ag_b} = (\bar\phi^T)_{g_ag_b} = \phi_{g_ag_b}$.  To make the link to group field theories more transparent, we shall define a related field: $\phi:\Zl_N\times \Zl_N \rightarrow \Cl$ and a operation $\dagger$ on the space of fields such that $\phi^\dagger = \phi$ where $\phi^{\dagger}(g_a,g_b) = \bar\phi(g_b,g_a)$.   Then, the action is usually taken to be:
\be
\begin{array}{rcl}
\cs[\phi] &=& \dsty\sum_{g_a,g_b}\phi(g_a,g_b)\;\bar\phi(g_a,g_b)- \frac{\lambda}{2\sqrt{N}}\sum_{g_a,g_b,g_c} \phi(g_a,g_b)\,\phi(g_b,g_c)\,\phi(g_c,g_a) \\[0.5cm]
&&\dsty \hspace{5cm}-  \frac{\lambda}{2\sqrt{N}}\sum_{g_a,g_b,g_c} \bar\phi(g_a,g_b)\,\bar\phi(g_b,g_c)\,\bar\phi(g_c,g_a) \\[0.6cm]

&=&\dsty  tr_N(\phi^2) - \frac{\lambda}{\sqrt{N}} tr_N(\phi^3).
\end{array}
\ee
where $tr_N$ is the usual trace over $N\times N$ matrices and we have used hermiticity to simplify the form of the action.
Let us examine the free energy:
\be
\cf = \log\cz = \int \cd\phi\;  e^{-S[\phi]} = \sum_{\calr}\sum_{\Gamma_\calr} \frac{1}{|Aut[\Gamma_\calr]|}\cz_{\Gamma_\calr} 
\ee
where:
\be
\int \cd \phi =\dsty\int \prod_{g_a} d\phi_{g_ag_a} \prod_{g_a,g_b: g_a < g_b} d\Fr\fe(\phi_{g_ag_b})\; d\Fraki\fm(\phi_{g_ag_b})
\quad\quad\textrm{and}\quad\quad
 \cz_{\Delta_\calr} =   \lambda^{\#f} N^{\chi(\calr)}  
\ee
The perturbative expansion of the free energy generates a sum over 2d orientable connected triangulations.  We shall say more about the Feynmanology in a moment. We are in a quantum mechanical setting so we can define the measure precisely by taking the Lebesgue measure on each of the matrix components. Since orientable Riemann surfaces are classified by a single parameter, the Euler characteristic $\chi(\calr) = \chi(\Gamma_\calr)$, we can consider the expansion as a sum over Riemann surfaces $\calr$ and tri-valent cellular decompositions $\Gamma_\calr$. Thus, the matrix model naturally encodes a sum over topologies.  We construct the Feynman amplitudes as follows. Due to the matrix structure of the fields, the strand diagrams associated to the propagator and vertex operator are, see Fig. \ref{2dfat}:
\begin{figure}[h]
\centering
\includegraphics[width = 15cm]{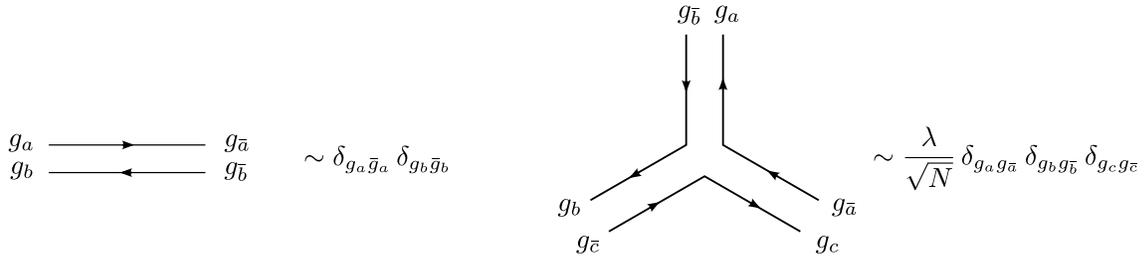}
\caption{\label{2dfat} Diagrammatics for $\phi^3$ hermitian matrix model.}
\end{figure}

\noindent It is easy to see that these fat graphs $s(\Gamma_\calr)$ are topologically dual to 2d triangulations $\Delta_\calr$.  The vertex operator generates the triangles and the propagator glues them pairwise along edges.  From here,  one can calculate the amplitude associated to a given triangulation in the sum.  There is a factor $\lambda/\sqrt{N}$ for every vertex $v\in \Gamma_\calr$ and for every face of $\Gamma_\calr$ there is a redundant sum which contributes a factor $N$.  Thus, the contribution of the $\Gamma_\calr$ to the free energy is:
\be
\cz_{\Gamma_\calr} = \lambda^{\# v} N^{\#f - \frac12\#v } = \lambda^{\# v} N^{\#v -\#e +\#f } = \lambda^{\# f} N^{\chi(\calr) } =  \lambda^{\# f} N^{2-2g_\calr}
\ee
where $v$, $e$ and $f$ are the vertices, edges and faces of $\Gamma_\calr$, respectively;  which satisfy $3\#v= 2\#e$; and finally $\chi(\calr)$ is the Euler character which satisfies $\chi(\calr) = \chi(\Gamma_\calr) = \#v -\#e +\#f = 2 - 2g_\calr$, where $g_\calr$ is the genus of the Riemann surface.

This amplitude looks familiar, however, from 2d quantum gravity.  In two dimensions, the Einstein-Hilbert action with Riemannian metric is integrable:
\be
\cs_{\calr}[g] = -\frac{1}{16\pi G}\int_\calr d^2x \sqrt{g}(R[g] -2\Lambda) = -\alpha\chi(\calr) + \beta A_\calr
\ee
where $g$, $R[g]$ are the metric and Ricci scalar; $G$ and $\Lambda$ are the gravitational and cosmological constants; while $\alpha = 1/4G$ and $\beta = \Lambda/8\pi G$.  Thus, on a fixed manifold, 2d gravity is trivial.  In the quantum case, however, it is no longer trivial, since large quantum fluctuations may change the genus of the surface and on higher genus surfaces, there are non-trivial topological sectors. Thus, one should define the partition function for 2d quantum gravity as:
\be
\cz_{2d} = \sum_{\calr} \int_X \cd g\; e^{-\cs_\calr[g]}, \quad\quad \textrm{where}\quad\quad X = Riem[\calr]/Dif\!f[\calr]
\ee
 with $Riem[\calr]/Dif\!f[\calr]$ the space of Riemannian metrics modulo diffeomorphisms. If one considers the matrix model to generate triangulations with every triangle assigned unit area, then $A_{\Delta_\calr} = \#f$.  Thus, equating:
 \be
  N = e^{\alpha}, \quad\quad \lambda = e^{-\beta},  \quad\quad\int_X \cd g = \sum_{\Delta_\calr}\frac{1}{|Aut[\Gamma_\calr]|}, 
 \ee
 we see that the free energy of the matrix model gives the partition for 2d gravity:
 \be
 \cf = \cz_{2d}
 \ee
 Furthermore, one can see immediately from the topological expansion that the $2$-sphere dominates in the large-$N$ limit (also called the planar limit for that reason):
 \be
 \cf \;\stackrel{N\rightarrow \infty}{\sim}\; \cz_{2d,S^2} = \sum_{\Gamma_{S^2}}\frac{1}{|Aut[\Gamma_{S^2}]|}\cz_{\Gamma_{S^2}}.
 \ee
  More interesting is the so-called double scaling limit.  It turns out that for large numbers of vertices: 
\be
  \frac{1}{N^2} \cz_{\Gamma_{S^2}}\; \stackrel{\#v \;\textrm{large}}{\sim} \; (\lambda - \lambda_c)^{2 - \gamma}\q\q \textrm{where} \q\q \gamma >2
  \ee
Thus, it diverges for some critical coupling and we can extract a continuum limit by tuning $\lambda\rightarrow \lambda_c$.     We refer the reader to \cite{difran} for details.

 The presence/absence of critical behaviour is one of the most striking and important features of any matrix model.  
In fact, for the one-hermitian matrix model, we can generalize to other potentials (eg. n-valent graphs dual to n-gons). The continuum limit still exists and describes pure gravity once again \cite{kazpot}.

Furthermore, this model is an example of a class of matrix models which have been solved exactly, in this case, using the hermiticity of the matrix to recast the problem in terms of its eigenvalues.   As a final point, which is of particular interest when studying quantum gravity regimes, one can examine the Schwinger-Dyson equations for this model, which generate a loop algebra, which is in turn a part of the Virasoro algebra \cite{loop}.

\subsubsection{2d Ising and Potts models}

The Ising model has a natural realization as a two-matrix model, where the two matrices represent the $\pm$ states of an Ising spin system.  Once again, we can think of it as a field $\phi:\Zl_N\times\Zl_N \rightarrow \Cl$ which also carries the defining $\Zl_2$ representation.  Thus, $\phi = \phi_{h_i}$, where $h_i\in \Zl_2$.  The dynamics are specified by the action:
\be\label{ising01}
S[\phi] = \sum_{h_i,h_j\in\Zl_2} tr_N (\phi_{h_i}\, K^{h_ih_j}\, \phi_{h_j}) + \frac{\lambda}{N} \sum_{h_i\in\Zl_2} tr_N(\phi_{h_i}^4)
\quad\quad \textrm{and}\quad\quad 
K^{h_ih_j}  = 
	\left(\begin{array}{rr}
	1 & -c\\[0.1cm]
	-c & 1
	\end{array}
	\right)
\ee
Setting $c = e^{-2\beta}$, we find that the free energy takes the form:
\be\label{ising02}
\cf = \sum_\calr \sum_{\Gamma_\calr} \frac{1}{|sym[\Gamma_{\calr}]|} \cz_{\Gamma_\calr} \quad\quad \textrm{where}\quad\quad \cz_{\Gamma_\calr} = \tilde \lambda^{\#v} N^{\chi(\calr)} \exp\Bigg(\beta \sum_{e\in \Gamma_\calr} h_{s(e)} h_{t(e)}^{-1}\Bigg)
\ee
and $\tilde \lambda = \frac{\lambda c}{(1 - c^2)^2}$. We see that the perturbative expansion generates 4-valent Feynman graphs which are dual to quadrangulations of Riemann surfaces.  The action contains two types of vertex interaction, representing the two spin states of the system, while the propagators couple the vertices in a nearest neighbour fashion.  Thus, we see that the amplitudes are that of an Ising model on the lattice $\Gamma_\calr$.

It turns out that the Ising model is much easier to solve when summed over random lattices.  Moreover, it is solvable even in the presence of a non-zero magnetic field, see \cite{ising}.  This ease stems from the extra symmetry gained by coupling to gravity, since the difficulties inherent in dealing with a fixed lattice are integrated out.

This form of matrix model can be extended to the $q$-state Potts model rather easily.   Now take $q$ $N\times N$ Hermitian matrices, or equivalently a field $\phi:\Zl_N\times\Zl_N\rightarrow \Cl$ which also carries the defining representation of $\Zl_q$, so that $\phi = \phi_{h_i}$ where $h_i\in\Zl_q$.  The action is a simple generalization of the Ising form:
\be\label{ising03}
S[\phi] = \sum_{h_i,h_j\in\Zl_q} tr_N (\phi_{h_i}\, K^{h_ih_j}\, \phi_{h_j}) - \frac{\lambda}{N} \sum_{h_i\in\Zl_q} tr_N(\phi_{h_i}^4)
\ee
and
\be 
K^{h_ih_j}  = 
	\left(\begin{array}{rrrr}
	1 & -c_q & \dots & -c_q\\[0.1cm]
	-c_q & 1 & \dots & -c_q\\[0.1cm]
	\vdots &\vdots&& \vdots\\[0.1cm]
	-c_q & -c_q&\dots& 1
	\end{array}
	\right)\q\q \textrm{where}\q\q c_q = \frac{1}{e^{2\beta} + q - 2}.
\ee
While one can still examine such features as critical behaviour, these models have not been solved exactly although there are arguments to suggest that this is possible \cite{kazpotts}.

\subsection{More general GFTs}

\subsubsection{n-dimensional topological models: Boulatov-Ooguri models}

Currently, one possesses a less well developed arsenal with which to attack higher-dimensional models.  All the same, to start off, we can check to see what classes of spin foam amplitudes can be reproduced and to what dynamical systems they can be associated.  For the topological models set out in Section \ref{top}, we can proceed directly to the GFT formalism.
We consider a field $\phi:G^{\times n}\rightarrow \Cl$. The action in this case is:
\be
\begin{array}{rcl}
\cs[\phi,\bar\phi] &=&\dsty \int_{G^{\times 2n}}\left[ \prod_{a=1}^n dg_a\; d\hat g_a\right] \phi(g_1, \dots, g_n) \left[\prod_{a = 1}^n \delta(g_a \hat g_a^{-1})\right] \bar\phi(\hat g_1,\dots,\hat g_n)\\[0.5cm]
 &&\dsty- \lambda \int_{G^{\times n(n+1)}} \left[\prod_{a,b = 1:b\neq a}^{n+1} dg_{a,b}\right] \left[\prod_{a = 1}^{n+1} \phi(g_{a,a-1},\dots,g_{a,a-n})\right]\left[ \prod_{a,b = 1:b\neq a}^{n+1} \delta(g_{a,b}^{\phantom{-1}}\,g_{b,a}^{-1})\right].\\[0.5cm]
 \end{array}
\ee
where unlike matrix models, it is conventional to use a normalized measure on the group (these normalized integrals are replaced by normalized sums in the finite group case).
The key ingredient is that the fields possess a symmetry alluded to earlier:
\be
\phi (g_1,\dots, g_n) = \phi(g_1 g,\dots,g_n g) \q\q \textrm{for any} \q\q g\in G.
\ee
We enforce this symmetry by explicit use of a projector onto the space of invariant fields, which results in the following propagator and vertex operator:
\be
\cp =\int_G dg \left[\prod_{a = 1}^n \delta(g_a g\hat g_a^{-1})\right] , \q\q \cv = \int_{G^{\times n+1}} \left[\prod_{a=1 }^{n+1}dg_a\right]\left[ \prod_{a,b = 1:b\neq a}^{n+1} \delta(g_{a,b}^{\phantom{-1}}\;g_a^{-1}\;g_b^{\phantom{ -}}g_{b,a}^{-1})\right]
\ee
By construction, the amplitude for a given Feynman graph:
\be\label{bf04}
\cz_{\Gamma, BF} = \lambda^{\#v}\int \prod_{e\in\Gamma}dg_e \prod_{f\in\Gamma} \delta(G_f)\q\q\textrm{where}\q\q G_f = \prod_{e\in \partial f} g_e^{o(e,f)}
\ee
Thus, we see that we have generated the amplitude associated to an $n$-dimensional BF theory \eqref{z4}.  This class of models is to some extent the starting point for the analysis of higher-dimensional GFTs.  For $G$ either a finite or  Lie group, the Feynman amplitudes have been well studied and their properties documented \cite{gurautop, cell}.  Various aspects have been analysed, such as the insertion of topological matter defects \cite{topmat}, fermionic matter coupling \cite{ferm}, the first instantonic solutions \cite{inst} have been found and analysed, a dual representation has been proposed \cite{dual} along with a first class of radiative corrections \cite{rad}.   Since these theories are background independent, it is natural to expect some trace of diffeomorphism symmetry \cite{diff} and the first steps in the implementation of mean field theory techniques have been completed \cite{mean}.  

Most outstandingly, it appears we can still capture, in these tensor models,  certain aspects of the topological expansion we saw for matrix models.  Indeed, at leading order, for the coloured models \cite{colour}, the $n$-sphere dominates \cite{gurautop, cell}.  Although, these are all interesting topics and it would be wonderful to dwell on them in more detail, our aim is to give a more basic outline of how GFTs can capture interesting dynamical systems.  With this in mind, let us pass to the higher-dimensional nearest neighbour models.

\subsubsection{$n$-dimensional Ising/Potts models and more general nearest-neighbour models}

There seems to be a rather general algorithm to generate Ising/Potts models on higher-dimensional random lattices.  This is simply to copy the 2d case.  Let us consider $q$ complex fields $\phi_{h}:G^{\times n}\rightarrow \Cl$, where $h\in\Zl_q$ and with the action as before:
\be
\begin{array}{l}
\cs[\phi,\bar\phi] =\dsty \int_{G^{\times 2n}}\left[ \prod_{a=1}^n dg_a\; d\hat g_a\right] \sum_{h_i,h_j\in\Zl_q}\phi_{h_i}(g_1, \dots, g_n)\; K^{h_ih_j} \left[\prod_{a = 1}^n \delta(g_a \hat g_a^{-1})\right] \bar\phi_{h_j}(\hat g_1,\dots,\hat g_n)\\[0.5cm]
 \phantom{xx}\dsty- \lambda \sum_{h_i\in\Zl_q} \int_{G^{\times n(n+1)}} \left[\prod_{a,b = 1:b\neq a}^{n+1} dg_{a,b}\right] \left[\prod_{a = 1}^{n+1} \phi_{h_i}(g_{a,a-1},\dots,g_{a,a-n})\right]\left[ \prod_{a,b = 1:b\neq a}^{n+1} \delta(g_{a,b}^{\phantom{-1}}\,g_{b,a}^{-1})\right].\\[0.5cm]
 \end{array}
\ee
In this case, one gets a Potts model coupled to the $BF$ theory outlined in the previous section:
\be\label{nising02}
\cz_\Gamma = \cz_{\Gamma,BF} \left[\prod_{v\in\Gamma}\sum_{h_v\in\Zl_q}\right]\exp\Bigg(\beta \sum_{e\in \Gamma} h_{s(e)} h_{t(e)}^{-1}\Bigg),
\ee
where the coupling constant occurring in \eqref{nising02} is the $\tilde\lambda$ of equation \eqref{ising02}.
In fact, we can generalize to even more general nearest-neighbour amplitudes just by altering the propagator $K^{h_ih_j}$.  Consider $H$ a finite (or compact) group.  Essentially, choosing:
\be
K^{h_ih_j} = \left(w_e(h_ih_j^{-1})\right)^{-1}
\ee
is enough to generate amplitudes of the form:
\be
\cz_\Gamma = \cz_{\Gamma,BF} \left[\prod_{v\in\Gamma}\sum_{h_v\in H}\right] \prod_{e\in \Gamma} w_e(h_{s(e)}h_{t(e)}^{-1})
\ee
Of course, for these nearest-neighbour interactions, the matter degrees of freedom are always coupled to some other theory, such as a $BF$ theory in this case. This is exactly analogous to the 2d scenario.

\subsubsection{$n$-dimensional Yang-Mills and more gauge theory models}

A different approach must be taken for the models of Section \ref{gauge}. It is more tricky to generalize to amplitudes of the form:
\be
\cz_\Gamma =  \lambda^{\#v}\int \prod_{e\in\Gamma}dg_e \prod_{f\in\Gamma} w_f(G_f) \q\q\textrm{where} \q\q G_f = \prod_{e\in \partial f} g_e^{o(e,f)}.
\ee
These are very much in the $BF$-type \eqref{bf04}, but they need not be topological.  As mentioned before, they may be viewed as a modification of the topological amplitude.  We have seen that lattice Yang-Mills amplitudes are in this form.  But the GFT formalism is not at first sight the most natural setting for these amplitudes.  This stems from the fact that we do not have control over face amplitudes at the outset.  The faces $f\in\Gamma$ arise after gluing of the interaction vertices. We do have control, however, over the amplitude assigned to part of the face. This part is known as a wedge in the spin foam literature \cite{wedge} and we shall denote it $f_v$.  Generically, the best we can do is to specify the vertex operator to be:
\be
\cv = \int_{G^{\times n+1}} \left[\prod_{a=1 }^{n+1}dg_a\right]\left[ \prod_{a,b = 1:b\neq a}^{n+1} w_{f_v}(g_{a,b}^{\phantom{-1}}\;g_a^{-1}\;g_b^{\phantom{ -}}g_{b,a}^{-1})\right]
\ee  
Then, the amplitudes turn to be:
\be
\cz_\Gamma = \lambda^{\#v}\int \prod_{e\in\Gamma}dg_e \prod_{f\in\Gamma} w_{f,\#}(G_f)\q\q\textrm{where}\q\q G_f = \prod_{e\in \partial f} g_e^{o(e,f)}
\ee
The function $w_{f,\#}(g)$ is defined in the following fashion:
\be
w_{f,\#}(g)  = \circ^{\#v\subset \partial f} w_{f_v} (g) \q\q \textrm{and}\q\q w_{f_v}\circ w_{f_v} (g) = \int_{G}dh\; w_{f_v}(gh^{-1})\;w_{f_v}(h).
\ee
In simple terms, not all the face amplitudes turn out to be the same.  But there is an obvious case when they do all coincide, that is, when $w_{f_v}$ is stable under group convolution.  A nice way of ensuring this is to attempt to find the 2d fixed point face amplitude (no matter what the dimension of the GFT in question).   This is still a highly non-trivial problem and we could ask for a slightly more lenient property, say, that the functional form of the wedge amplitude is stable under convolution.  To take Yang-Mills as an example, the Wilson amplitude does not suffice since:
\be
w_{f_v}(g) = \exp\left(-\alpha\;\Fr\fe\chi^\rho(g)\right) 
\ee
does not maintain the same functional form under convolution of the wedges.  But it is well known that the heat kernel amplitude is  stable under convolution (at the level of each face):
\be
w_{f_v}(g) = \sum_{\rho\in G}d_\rho\;\chi^\rho(g)\; \exp\left(-\alpha\;C^\rho\right) \implies w_{f,\#}(g) = \sum_{\rho\in G}d_\rho\;\chi^\rho(g)\; \exp\left(-(\#f_v\times\alpha)\;C^\rho\right)
\ee
where $d_\rho$, $C^{\rho}$, $\chi^\rho$ are the dimension, quadratic Casimir and character of the representation of $G$ labelled by $\rho$; $\alpha$ is the coupling constant and $\#f_v$ is the number of wedges in $f$.   Although the weight maintains the same functional form under convolution, one acquires a $\#f_v$ dependence for each face and $\#f_v$ differs for each face.  This is fine in the pure gauge theory setting since the heat kernel amplitude generically comes with a factor of the area of the face.  If we choose all wedges to have unit area, then $\#f_v$ is exactly the area of the face that we wish to pick up.   Of course, if one wishes to couple this theory to gravity at the GFT level, then one would like this area factor to be generated dynamically from the gravity side of the theory.  As yet, there are no proposals for such a mechanism existing in the literature.   Perhaps what is most interesting at this stage, is that knowing about the 2d theory can guide us, at least initially, in setting out the $n$-dimensional theory.  The action for the gauge model above is:
\be\label{plaq}
\begin{array}{rcl}
\cs[\phi,\bar\phi] &=&\dsty \int_{G^{\times 2n}}\left[ \prod_{a=1}^n dg_a\; d\hat g_a\right] \phi(g_1, \dots, g_n) \left[\prod_{a = 1}^n K(g_a \hat g_a^{-1})\right] \bar\phi(\hat g_1,\dots,\hat g_n)\\[0.5cm]
 &&\dsty- \lambda \int_{G^{\times n(n+1)}} \left[\prod_{a,b = 1:b\neq a}^{n+1} dg_{a,b}\right] \left[\prod_{a = 1}^{n+1} \phi(g_{a,a-1},\dots,g_{a,a-n})\right]\left[ \prod_{a,b = 1:b\neq a}^{n+1} \delta(g_{a,b}^{\phantom{-1}}\,g_{b,a}^{-1})\right].
 \end{array}
\ee
where $K\circ w_{f_v}(g) = w_{f_v}\circ K(g) = \delta(g)$.  There is some freedom here, we could replace the $\delta$-functions in the vertex operator with $w_{f_v}$ instead and leave the propagator unchanged, but we prefer to act in this way due to its similarity with another class of plaquette theories.  These are modifications of $BF$-theory which can be viewed as tentative proposals for higher dimensional quantum gravity, the EPRL/FK group field theories.  These have a similar functional form to \eqref{plaq}, obtained by modifying the propagator.  Details of this construction can be found in \cite{fra} and references therein.

\section{Non--local spin foams} \label{nonlocal}

One of the main open problems in quantum gravity is to study the large scale behaviour of the models. For this renormalization group methods could be clearly useful. However the application of renormalization group methods  to spin foam models  currently discussed as quantum gravity candidates \cite{barrett_crane,eprl} is hindered not only by many conceptual issues but also by the tremendous complicated amplitudes emerging from these models.  Here `toy spin foam models' could help to develop new techniques and to investigate the statistical field aspects of these models, some of which might be independent of the specifics of the amplitudes. For instance the question whether and how to  sum over topologies of spin foams might not depend on the chosen group underlying the model. Also note that there are a number of quantum gravity approaches which rather work with quite simple (vertex) amplitudes \cite{z2gravity,dyntriang,causal}, in particular \cite{causal} in which the question of the large scale limit can be addressed. The models proposed in this work can be seen as small spin cut--off models of the full theory. The hope is that for these models one can probe the many particle (and small spin) regime which is in contrast to the very few particle and large spin regime considered up to now.

There has been some very interesting work exploring coarse graining in the spin foam context \cite{fotini,oeckl}. Also the related concept of tensor networks \cite{tensorn}, a generalization of spin nets introduced here, has been developed as a tool for coarse graining (to specify topological order in condensed matter systems).  Mostly these however concentrate on frameworks and truncations in which the local form of the spin foams does not change. One might however need to accommodate also for non--local\footnote{These couplings are in general exponentially suppressed with respect to some non--locality parameter, like distance or size of the Wilson loops. In this case one would still speak of a local theory.} couplings, in particular if one attempts to regain diffeomorphism symmetry, which is broken by the discretizations employed by many quantum gravity models \cite{perfect,coarse1}.

Indeed applying block spin transformations to a non--topological lattice gauge system one can expect to obtain a partition function of the form
\ba\label{r1}
Z&=&  \sum_{g_e}  \prod_{M=1}^\infty  \prod_{l_1, \ldots, l_M} w_{l_1,\ldots l_M} (h_{l_1},\ldots,h_{l_M})
\ea
where $h_{l_i}$ are holonomies around loops (that is around plaquettes but also around other surfaces made of several plaquettes) and $w_{l_1,\ldots l_M}$ is a class function of its arguments, describing possible couplings between (Wilson) loops. A character expansion would not only introduce representation labels on the basic plaquettes but also on all the other surfaces encircled by loops appearing in (\ref{r1}). The resulting structure is akin to a two--dimensional generalization of a graph. Graphs would appear as effective descriptions  for the coarse graining of the Ising like models discussed in section \ref{ising}. Such non--local `spin graphs' would be generalizations of the spin nets introduced there.   Similarly, graphs have been introduced in \cite{knopf} to accommodate non--local couplings and to describe phase transitions between geometric (i.e a phase where for instance a space time dimension can be defined) and non--geometric phases. We leave the exploration of the ensuing structures for future work.

\section{Outlook}\label{out}

In this work we discussed several concepts and tools which arose in the spin foam, loop quantum gravity and group field theory approach to quantum gravity and applied these to finite groups. We encountered different classes of theories, one is the well known example of Yang Mills like theories another are topological $BF$ theories. The latter are also well known in condensed matter and quantum computing. A third class of theories are the constrained models discussed in section \ref{constra} which mimic the construction of the gravitational models. These are only applicable for non--Abelian groups as for Abelian groups the invariant Hilbert space associated to the edges is always one--dimensional and cannot be further restricted. We plan to study these theories in more detail in the future, in particular the symmetries and the associated transfer or constraint operators.  The relative simplicity of these models (compared to the full theory) allows for the prospect of a complete classification of the choices for the edge projectors and hence of the different constrained models. For the study of the translation symmetries in the non--Abelian models it might be fruitful to employ a non--commutative Fourier transform \cite{nc,dual,diff}. This would define an alternative dual for the non--Abelian models, in which the edge projectors carry delta function factors, however defined on a non--commutative space.

We also mentioned the possibilities to obtain gravity like models by breaking down the translation symmetries of $BF$ theory. This could be particularly promising in the canonical formalism where one can be guided by the form of the Hamiltonian constraints for gravity. This strategy is also available for the Abelian groups. In particular for $\Zl_2$ it is in principle possible to parametrize all possible Hamiltonians or partition functions and to study the  associated symmetry content. See also the universality result in \cite{briegel}. Hence there might be a definitive answer whether gravity like models exists, i.e. $4D$ ($\Zl_2$) lattice models with translation symmetries based on the $4$--cells (or the dual vertices).

Finally we hope that these models can be helpful in order to develop techniques for coarse graining and renormalization of spin foam models and in group field theories. Here the connection to standard theories could be exploited and techniques can be taken over from the known examples and with adjustments be applied to quantum gravity models. To this end the finite group models could be an important link and provide a class of toy models on which ideas can be tested more easily. For instance the connection between (real space) coarse graining of spin foams and renormalization in group field theories, which generate spin foams as Feynman diagrams, could be explored more explicitly than in the (divergent) $SU(2)$--based models.

It will be particularly interesting to access the many--particle regime, for instance using Monte Carlo simulations, which for the full models is yet out of reach. In the ideal case it might be possible to explore the phase structure of the constrained theories and to study the symmetry content of these different phases.

\section*{Acknowledgements}

B.D. thanks Robert Raussendorf for discussions on topological models in quantum computing. The authors are thankful to Frank Hellmann for illuminating discussions.


\end{document}